# Emotional Expression in Persuasion by Quadruped Virtual Agents: Toward Cross-Species Design Patterns

Kaoru Sumi[1]* and Souki Osawa[1]
[1] Future University Hakodate, Hakodate, Japan
*Corresponding author: kaoru.sumi@acm.org

Abstract

Persuasive technologies increasingly employ virtual agents to influence human attitudes and behaviors. While prior research has largely focused on humanoid agents, the persuasive design of non-humanoid agents, particularly quadruped and animal-like agents, remains underexplored. Pet-like agents may evoke intuitive and affective responses, but it is still unclear whether emotional expression functions consistently across different animal species and whether species-specific motion is necessary for effective persuasion.

This study investigates the role of emotional expression in persuasion by quadruped virtual agents from a cross-species design perspective. We developed three virtual agents representing a dog, cat, and horse, and compared three behavioral conditions: species-specific behavior, shared behavior across species, and a bark-only baseline. Participants were presented with everyday behavioral tasks, including disposing of trash, feeding, and refraining from smartphone use. The agents attempted to influence participants through expressive behaviors, and the interaction was evaluated using multiple measures, including intention understanding, behavioral intention, actual behavior, psychological reactance, discomfort, familiarity, and agent acceptance.

The results showed that, in several task contexts, the bark-only baseline produced lower scores in intention understanding and several behavioral measures than the expressive behavior conditions. This suggests that persuasive behaviors incorporating emotional expression and attention-guiding cues improve users' interpretation of agent intention and support behavior change. In contrast, no consistent significant differences were found between the species-specific and shared behavior conditions. This indicates that faithful reproduction of animal-specific motion is not necessarily the primary determinant of persuasive effectiveness.

Psychological reactance and discomfort remained low across conditions, suggesting that persuasive attempts by quadruped agents did not strongly elicit negative responses. In addition, familiarity with the animal species was associated with actual behavior in some conditions, indicating that users' prior affinity toward a particular animal may also shape persuasive outcomes.

These findings suggest that persuasive effectiveness in quadruped virtual agents is driven more by functional cues such as emotional expression, attention guidance, and intention readability than by the accurate reproduction of species-specific behaviors. The study provides empirical evidence for cross-species generalizability in animal-like persuasive agents and offers a foundation for reusable design patterns in persuasive technology and Human–AI interaction for behavior change.

## 1. Introduction

Persuasive technologies aim to influence human attitudes and behaviors through interactive systems, feedback, social cues, and other forms of human–computer interaction. Since the foundational work on computers as persuasive and social actors, research has increasingly explored how digital systems can support behavior change in domains such as health promotion, education, environmental sustainability, and everyday habit formation (Reeves and Nass, 1996; Fogg, 2003; Oinas-Kukkonen and Harjumaa, 2009). In recent years, virtual agents and embodied interactive systems have become particularly important in persuasive technology because they can communicate intentions, provide social feedback, and evoke affective responses through verbal and nonverbal behaviors.

Most research on persuasive agents, however, has focused on humanoid agents, conversational agents, or socially interactive robots. These agents often rely on human-like speech, facial expressions, gestures, and social roles to influence users. In contrast, non-humanoid agents, especially animal-like and pet-type agents, remain comparatively underexplored as persuasive interfaces. This is an important gap because pet-type agents can evoke familiarity, empathy, care, and intuitive interpretation of nonverbal behavior. Unlike humanoid agents, animal-like agents may encourage users to respond not through explicit instruction or authority, but through affective cues such as apparent sadness, neediness, attention seeking, or vulnerability.

This perspective is closely related to what we refer to here as Persuasive and Affective Human–AI Interaction (PAHAI), which examines how AI agents can support human understanding, engagement, attitude change, and behavior change through affective, embodied, and socially meaningful interaction. In PAHAI, persuasion is not limited to rational information delivery or explicit recommendation. Rather, it can emerge from emotional expression, bodily movement, social presence, perceived intention, and users' emotional interpretation of the agent. This perspective also connects to recent work on affective learning and serious games, where emotional, embodied, and interactive systems are regarded as important mechanisms for supporting engagement, reflection, and behavior change (Sumi, 2026). Although the present study does not directly evaluate learning outcomes, it extends this broader perspective by examining how animal-like virtual agents can use emotional expression and nonverbal behavior to guide everyday actions.

Prior work has shown that emotional expression can play an important role in persuasion by pet-type agents. Harada and Sumi (2024) investigated persuasive technology through behavior and emotion with a dog-like pet-type artifact in mixed reality and showed that emotional expressions such as sadness and confusion could encourage users to perform a target behavior, whereas expressions such as joy and anger could be less effective or lead to misinterpretation. Subsequent work further examined how emotional expressions interact with behavioral cues such as attention calling, guiding, and pointing in mixed reality contexts (Sumi and Harada, 2025). These studies suggest that persuasive effects in pet-type agents depend not only on the presence of emotional expression, but also on whether the expressed emotion, movement, and behavioral context allow users to correctly infer the agent's intention.

However, an important question remains unresolved: are such persuasive effects specific to dog-like agents, or can they generalize across different animal species? Dogs have a distinctive history of social communication with humans, and users may readily interpret their gaze, vocalizations, and body movements as meaningful. In contrast, users may have different or less consistent expectations about the emotional and communicative behaviors of other animals, such as cats or horses. Therefore, it remains unclear whether persuasive behavior design should faithfully reproduce species-specific motion, or whether persuasive effectiveness can be achieved through species-independent functional cues, such as emotional expression, attention guidance, gaze alternation, and intention readability.

To address this question, the present study investigates persuasive behaviors in quadruped virtual agents from a cross-species design perspective. We developed virtual agents representing a dog, a cat, and a horse, and compared three behavioral conditions: species-specific behavior, shared behavior across species, and a bark-only baseline with a visible dog agent. The agents attempted to influence participants in everyday behavioral tasks, including disposing of trash, feeding, and refraining from smartphone use. Persuasive effectiveness was evaluated using multiple measures, including intention understanding, behavioral intention, actual behavior, psychological reactance, discomfort, familiarity, and agent acceptance.

The main contribution of this study is not merely the comparison of dog, cat, and horse agents. Rather, this study empirically examines whether persuasive behaviors in quadruped virtual agents can be abstracted as species-independent functional design elements. By comparing species-specific and shared behaviors across multiple animal-like agents, the study investigates whether persuasive effectiveness depends on faithful reproduction of animal-specific motion or on more general affective and communicative cues. In doing so, this work aims to provide initial empirical evidence for cross-species generalizability in animal-like persuasive agents and to contribute to reusable design patterns for persuasive and affective Human–AI Interaction.

This study makes three contributions. First, it extends research on pet-type persuasive agents from dog-like agents to multiple quadruped species. Second, it empirically examines whether persuasive effectiveness depends on species-specific motion or on species-independent functional cues such as emotional expression, attention guidance, and intention readability. Third, it provides design implications for standardizing persuasive behaviors in quadruped virtual agents, contributing to the development of non-humanoid affective persuasive interfaces for behavior change.

## 2. Related Work

This section reviews prior work that provides the theoretical and empirical background for the present study. First, we review persuasive technology and behavior change systems as the broader foundation for agent-mediated persuasion. Second, we discuss affective Human–AI interaction and the role of emotional and embodied cues in shaping user interpretation and engagement. Third, we examine pet-type and animal-like agents in mixed reality and related interactive environments. Fourth, we summarize prior work on emotional expression and behavioral cues in pet-type persuasive agents, including studies that directly informed the present research. Finally, we identify the remaining research gap concerning whether persuasive behaviors in animal-like agents are species-specific or can be generalized as species-independent functional design elements.

### 2.1 Persuasive Technology and Behavior Change

Persuasive technology refers to interactive systems designed to change users' attitudes, intentions, or behaviors without coercion. Foundational work on persuasive technology and media equation theory positioned computers not only as tools and media, but also as social actors capable of influencing human thought and behavior (Reeves and Nass, 1996; Fogg, 2003). This perspective is particularly important for agent-based persuasion because virtual agents and robots can present social cues, express intentions, and respond to users in ways that invite social interpretation.

Subsequent work on persuasive systems design has further emphasized that behavior change technologies should be understood not only in terms of information delivery, but also in terms of interaction design, credibility, social support, tailoring, and user engagement (Oinas-Kukkonen and Harjumaa, 2009). In this view, persuasion is not simply a matter of presenting users with rational arguments or instructions. Rather, it depends on how the system frames the target behavior, how it communicates with the user, and how the user perceives the system's role, intention, and social presence.

Virtual agents and embodied systems provide a particularly rich design space for persuasive technology. Unlike static interfaces, agents can use gaze, gesture, posture, movement, facial expression, vocalization, and spatial behavior to communicate intention and influence users' responses. These social and nonverbal cues can make the system appear more present, attentive, or relational, thereby shaping how users interpret its requests or recommendations. Prior research on persuasive robots has shown that both vocal and nonverbal cues can influence persuasive outcomes, and that physical or embodied cues may play an important role in users' compliance and interpretation of a robot's intention (Chidambaram et al., 2012).

Gaze and motion are especially important because they can communicate intention without explicit verbal explanation. Mutlu et al. (2009) showed that robot gaze behavior can shape participants' roles, attention, and participation in human–robot conversations, suggesting that gaze cues can structure social interaction. Takayama et al. (2011) further demonstrated that motion design based on animation principles can improve robot readability by helping users infer what a robot is doing or trying to do. These studies suggest that users interpret agents not only through what they say, but also through how they move, orient their body, direct attention, and prepare for action.

These findings indicate that persuasion by agents is not limited to verbal content. Instead, it can emerge from the coordination of multiple communicative channels, including movement, gaze, posture, sound, and spatial positioning. This is especially relevant for nonverbal or minimally verbal agents, where users must infer the agent's intention from its behavior rather than from explicit instruction. For such agents, the readability of intention becomes a central design issue.

However, much of this work has focused on humanoid agents, conversational systems, or robots designed for direct human–robot interaction. These systems often persuade through human-like communication, such as speech, facial expressions, gestures, or social roles. In contrast, animal-like and pet-type agents offer a different form of persuasive interaction. Rather than relying on linguistic explanation or human-like authority, they may influence users through affective and relational cues, such as apparent need, vulnerability, attention seeking, or dependency. This suggests that pet-type agents can be understood as a distinctive class of persuasive interfaces in which emotional expression and nonverbal behavior are central to intention interpretation and behavior change.

### 2.2 Affective Human–AI Interaction and PAHAI

Affective interaction has become an important perspective in Human–AI Interaction because users do not interpret artificial agents only as functional tools. They often respond to agents through emotional, social, and relational interpretations. Since the early development of affective computing, emotion has been regarded as an essential component of intelligent and human-centered interaction (Picard, 1997). Research on embodied conversational agents has also shown that affective and social cues are expressed not only through facial expressions, but also through gestures, gaze, posture, and the coordination of multiple modalities (Cassell et al., 2000; Niewiadomski and Pelachaud, 2007; Pelachaud, 2009). Similarly, studies on relational agents and sociable robots have emphasized that emotional expression, social behavior, and long-term relational cues can shape users' engagement, trust, and willingness to interact with artificial agents (Breazeal, 2003; Bickmore and Picard, 2005).

In this paper, we use the term Persuasive and Affective Human–AI Interaction (PAHAI) to refer to a research perspective that integrates persuasive technology, affective computing, embodied interaction, and Human–AI Interaction to examine how AI agents can support human understanding, engagement, attitude change, and behavior change through affective and socially meaningful interaction. In this view, persuasion is not limited to explicit recommendations, verbal instructions, or rational information. It can also emerge from emotional expression, bodily movement, perceived intention, social presence, and users' affective interpretation of the agent. This view extends the foundation of persuasive technology, where interactive systems are designed to change users' attitudes or behaviors without coercion (Fogg, 2003), by emphasizing the affective, embodied, and relational mechanisms through which persuasion may occur.

PAHAI is particularly relevant to non-humanoid and animal-like agents because these agents may not rely on human language or human facial expressions in the same way as humanoid agents. Instead, they communicate through nonverbal and affective cues, such as gaze direction, head movement, body orientation, approach behavior, vocalization, and apparent emotional state. Users may interpret such cues as signs of need, hesitation, sadness, anxiety, friendliness, or dependency. These interpretations can motivate supportive or caring behavior, especially when the agent appears vulnerable or in need of assistance. This interpretation is also consistent with the concept of weak robots, which suggests that a robot's apparent vulnerability or incompleteness can invite human assistance and create relational forms of interaction (Okada, 2023). In the context of pet-type persuasive agents, such vulnerability cues may function as affective triggers that encourage users to support the agent rather than simply obey it.

This perspective also connects to research on affective learning and serious games, where emotional engagement, embodied experience, and interactive feedback are considered important mechanisms for supporting motivation, reflection, and behavioral change (Sumi, 2026). Although the present study does not directly evaluate learning outcomes, it shares the same broader concern: how affective and embodied interaction with artificial agents can shape users' interpretation and action. In this sense, quadruped virtual agents can be regarded as affective persuasive interfaces that use emotional expression and nonverbal behavior to support everyday behavior change.

The present study contributes to PAHAI by examining whether persuasive behaviors in animal-like virtual agents can be designed around species-independent functional cues.
If users can understand an agent's intention and respond behaviorally even when the agent is not humanoid and even when the animal species differs, this suggests that affective persuasive interaction may be designed at a more abstract level than species-specific realism. Such a finding would extend PAHAI beyond humanoid agents and conversational systems toward a broader design space of non-humanoid, embodied, and affective AI agents.

### 2.3 Pet-Type and Animal-Like Agents: Controllability, Presence, and Mixed Reality

Pet-type and animal-like agents have been explored as socially acceptable and emotionally engaging forms of artificial agents. Unlike humanoid agents, which often evoke expectations of human-like intelligence, language, and social competence, pet-type agents can invite a different mode of interaction based on familiarity, care, playfulness, and affective attachment. Robotic pets such as AIBO, PARO, and other companion robots have been studied in contexts such as entertainment, therapy, elderly care, and social support, showing that animal-like forms can encourage interaction and emotional engagement without requiring human-like appearance or conversation (Fujita, 2001; Wada and Shibata, 2007).

The appeal of pet-type agents is partly grounded in the psychological and social effects associated with human–animal interaction. Animal-assisted interventions have been discussed as a way to support emotional well-being, social engagement, and therapeutic interaction (Kruger and Serpell, 2010; Fine, 2010). However, real animals are not always suitable for experimental or applied settings because their behavior cannot be fully controlled, and their use may involve constraints related to allergies, hygiene, safety, animal welfare, and individual variability. Pet-type robots and virtual animals therefore offer a distinctive advantage: designers can systematically control the agent's appearance, movement, emotional expression, timing, spatial behavior, and response patterns while preserving animal-like familiarity and affective appeal. This controllability makes it possible not only to design gentle interventions for everyday behavior change, but also to decompose the mechanisms of persuasion by comparing how different cues influence users' interpretation and action.

Animal-like agents are particularly relevant to persuasive interaction because people are accustomed to interpreting animal behavior through nonverbal cues. Dogs, for example, communicate with humans through gaze, body orientation, movement, vocalization, and approach behavior. Studies of human–dog communication have shown that dogs can respond to human gaze and pointing, use gaze alternation and body movement to communicate intention, and be interpreted by humans as intentional or communicative in everyday contexts (Miklósi et al., 1998; Miklósi et al., 2000; Soproni et al., 2001; Lakatos et al., 2012; Bradshaw and Rooney, 2016). These findings suggest that animal-like agents may be able to convey intention and invite human response through relatively simple nonverbal behaviors. This idea is also supported by research on hearing-dog-inspired robots, which suggested that gaze direction and head movement can help users infer an agent's communicative intention (Koay et al., 2013).

This point is important for persuasive technology because animal-like agents may influence users without using explicit verbal commands. A pet-type agent that looks at an object, moves toward it, alternates its gaze between the user and the object, or appears sad or troubled may lead users to infer that the agent wants help. Such interaction differs from conventional instruction-based persuasion. Rather than telling users what to do, the agent creates a situation in which users interpret its state and voluntarily choose to act. This mechanism is closely related to the concept of weak robots, which suggests that an agent's apparent vulnerability, incompleteness, or dependence on others can invite human assistance and create a relational form of interaction (Okada, 2023). Studies on weak robots, such as the sociable trash box, have also shown that robots can elicit human assistance by displaying limited capability or neediness rather than by issuing explicit commands (Yamaji et al., 2011). In the context of pet-type persuasive agents, sadness, hesitation, or neediness may therefore function not as a sign of failure, but as an affective and relational cue that elicits supportive behavior. Together, these studies support the idea that pet-type persuasive agents may influence behavior through interpretable nonverbal cues, perceived vulnerability, and relational engagement rather than through explicit instruction.

Mixed reality and augmented reality technologies further expand the design space of pet-type and animal-like agents. In MR and AR environments, virtual agents can appear in the user's physical space and interact with real objects, places, and tasks. This makes it possible for animal-like agents to guide attention toward physical targets, indicate locations, or behave as if they share the same environment with the user. Prior work on virtual animals and AR companions has suggested that co-present animal-like agents can influence users' spatial behavior, awareness, attention, and sense of social presence (Norouzi et al., 2019; Norouzi, 2021). In this sense, MR provides a promising platform for persuasive animal-like agents because it allows affective and nonverbal cues to be grounded in the user's real environment.

Presence is also important for understanding persuasive interaction in MR. Social presence refers to the extent to which a mediated or virtual entity is experienced as being socially present and capable of interaction (Short et al., 1976; Biocca et al., 2003). More broadly, presence has been discussed as the perceptual experience of “being there” or the illusion that a mediated experience is not mediated (Lombard and Ditton, 1997; Lee, 2004). In MR environments, virtual agents can be spatially anchored in the user’s physical surroundings, which may strengthen the impression that the agent shares the same environment and is directing attention toward real objects or locations. This sense of co-presence can make nonverbal cues, such as gaze, pointing, approach behavior, and emotional expression, more meaningful for users.

Importantly, MR suggests that physical robotic embodiment is not the only way to create a sense of presence. In mixed reality, a virtual agent can be spatially anchored in the user’s physical environment and can direct its gaze, movement, and emotional expression toward real objects and locations. This allows users to experience the agent as being situated in the same space, even when the agent is not physically embodied as a robot. In this sense, MR provides a distinctive advantage for persuasive animal-like agents: it combines the controllability of virtual agents with the situatedness and social presence associated with embodied interaction. Designers can systematically manipulate the agent’s species, motion, emotional expression, timing, and spatial relationship to real objects while preserving the impression that the agent is co-present in the user’s environment.

At the same time, animal-like agents raise specific design challenges. Their persuasive effects may depend on how users interpret the agent’s species, appearance, motion, and emotional expression. For example, a behavior that is easily understood when performed by a dog may not be interpreted in the same way when performed by a cat, a horse, or another animal. Users may also differ in their familiarity with each animal species, which can affect whether they perceive the agent’s behavior as meaningful, friendly, confusing, or persuasive. Research on virtual animals has also pointed out that appearance and realism can influence users’ affective responses, including discomfort or uncanny impressions (Schwind et al., 2018).

Therefore, while pet-type and animal-like agents provide a promising alternative to humanoid persuasive agents, it remains necessary to clarify which design elements are species-specific and which can be generalized across animal-like forms. This issue is central to the present study, which compares dog, cat, and horse virtual agents to examine whether persuasive effects are tied to animal-specific motion or to more general functional cues.

### 2.4 Emotional Expression and Behavioral Cues in Pet-Type Persuasive Agents

Emotional expression has long been considered an important factor in persuasion and social influence. Studies on emotion in negotiation have shown that expressed emotions can shape how people interpret others' intentions and decide how to respond (Morris and Keltner, 2000). In particular, different emotions can have different effects depending on the interactional goal and context, suggesting that emotional expression does not have a uniform persuasive function (Morris and Keltner, 2000; Sinaceur and Tiedens, 2006). In Human–Agent Interaction, prior work has also suggested that a virtual agent's emotional expression and impression can influence users' willingness to accept a persuasive message or perform a requested action (Sumi and Nagata, 2010). These findings provide a basis for examining how emotional expressions in pet-type agents influence users' interpretation and behavior.

A direct line of prior work has examined how pet-type agents can persuade users through emotional expression and nonverbal behavior. Harada and Sumi (2024) investigated persuasive technology through behavior and emotion with a dog-like pet-type artifact in a mixed reality environment. Their study focused on a situation in which a virtual dog attempted to encourage users to throw away trash. Emotional expressions were designed by combining multiple components of canine expression, including head and neck posture, ear and tail movements, vocalization, and movement speed. Based on these components, six emotional expressions were implemented: neutral, sadness, anger, joy, surprise, and confusion. These emotional expressions were combined with two movement patterns that differed in the number and length of actions.

The results of Harada and Sumi (2024) showed that sadness and confusion were effective in motivating participants to perform the target behavior. Participants often interpreted these expressions as indicating that the dog was sad, troubled, or in need of help. These interpretations increased their willingness to throw away trash. In contrast, joy and anger were less effective in this context. Joy tended to be interpreted as a request for play rather than a request to throw away trash, while anger could make the dog appear frightening or difficult to approach. These findings are consistent with the broader view that the persuasive effect of emotion depends on how the emotion is interpreted in relation to the target behavior. An emotion that supports one persuasive goal may be inappropriate or counterproductive for another goal.

This finding is important because it shifts the design focus from emotional expression as a surface-level feature to emotional expression as a functional cue for intention interpretation. In pet-type persuasive agents, the goal is not simply to make the agent appear emotional. The agent's emotional state must be readable in relation to the target behavior. Sadness or confusion may invite supportive action when the target behavior can be interpreted as helping the agent. However, joy may redirect the user's interpretation toward play, and anger may generate avoidance rather than compliance. Therefore, emotional expressions must be selected and designed in relation to the behavioral goal and the user's expected interpretation.

Building on this line of work, Sumi and Harada (2025) further examined persuasive interactions with pet-type virtual agents by focusing on the combination of emotional expressions, behavioral actions, and context in mixed reality. Their study designed a dog-like virtual agent that combined emotional expressions, such as sadness, happiness, anger, and neutral states, with three types of behavioral actions: attention calling, guiding, and pointing. Attention calling involved approaching the user, barking, and making eye contact. Guiding involved moving toward a target location. Pointing involved alternating gaze or head orientation between an object and its destination. These behavior types are particularly relevant to pet-type agents because they allow the agent to communicate intention without relying on explicit verbal commands.

The study by Sumi and Harada (2025) extended the earlier work in two important ways. First, it examined multiple daily contexts, including pet-related tasks, non-pet-related tasks, limiting smartphone use, and emergency warning situations. This made it possible to analyze whether the same emotional expression had different persuasive effects depending on the context. Second, it compared mixed reality presentation with video-based presentation, thereby examining how the sense of co-presence and interactivity influences users' interpretation of the agent. The results suggested that persuasive effects depended on the combination of emotion, action, and context. Sadness combined with pointing or guiding was often perceived as supportive and persuasive, whereas anger could cause discomfort or become counterproductive depending on the situation. The study also suggested that mixed reality presentation can increase social presence and users' expectations for interactivity and responsiveness.

Together, these studies provide an important foundation for the present research. Harada and Sumi (2024) demonstrated that emotional expressions such as sadness and confusion can support persuasive behavior in a dog-like pet-type artifact when the expressed emotion is consistent with the target behavior. Sumi and Harada (2025) further showed that persuasive effectiveness depends on the integration of emotional expression, behavioral cues, contextual appropriateness, and the sense of presence created by mixed reality. These findings suggest that persuasive effectiveness in pet-type agents arises not from the mere presence of an agent or from emotional expression alone, but from the way users interpret the agent's intention, need, or emotional state through multimodal and situated cues.

However, these studies primarily focused on dog-like agents. Therefore, it remains unclear whether the observed persuasive effects are specific to canine communication or whether similar mechanisms can generalize to other quadruped animal-like agents. This distinction is important because a behavior that is easily interpreted in a dog may not be interpreted in the same way in a cat, a horse, or another animal. It is also unclear whether persuasive behavior design should rely on faithful reproduction of species-specific motion or whether more abstract functional cues, such as emotional expression, attention guidance, and pointing, can support presence and intention readability across species. The present study addresses this gap by comparing dog, cat, and horse virtual agents and by examining whether persuasive effectiveness depends on species-specific behavior or on species-independent functional design elements.

### 2.5 Research Gap and Contributions

The studies reviewed above suggest that pet-type and animal-like agents can provide a distinctive form of persuasive interaction. Unlike humanoid agents, they may influence users through affective and relational cues rather than explicit verbal instruction or authority. Prior work on pet-type persuasive agents has shown that emotional expression, movement, contextual appropriateness, and the sense of presence created by mixed reality are important for helping users interpret an agent's intention and respond behaviorally (Harada and Sumi, 2024; Sumi and Harada, 2025). These findings are also consistent with broader research on persuasive technology, affective interaction, weak robots, social presence, and animal-like agents in mixed reality environments.

However, three important gaps remain. First, previous studies have primarily focused on dog-like agents. Dogs are familiar social animals, and users may have shared expectations about canine gaze, vocalization, tail movement, and approach behavior. It is therefore unclear whether the persuasive mechanisms observed in dog-like agents are specific to canine communication or whether they can be generalized to other animal-like agents.

Second, previous work has not sufficiently distinguished between species-specific motion and more general functional cues. In animal-like agent design, it is often assumed that reproducing animal-specific behavior increases naturalness and effectiveness. However, for persuasive interaction, what may matter most is not whether the motion is faithful to a particular animal species, but whether users can read the agent's intention, emotional state, and behavioral goal. This distinction is important for design because species-specific animation requires separate modeling for each animal, whereas species-independent functional cues could support reusable behavior design across multiple agents.

Third, the relationship between affective persuasion and psychological burden remains underexplored in animal-like agents. Persuasive systems must not only promote target behaviors, but also avoid excessive discomfort, resistance, or psychological reactance. Psychological reactance is particularly important in persuasive interaction because users may resist or reject attempts to influence their behavior when they perceive the interaction as controlling or intrusive (Brehm, 1966). Pet-type agents may offer a gentle and socially acceptable form of persuasion, but it is still necessary to examine whether persuasive behaviors by animal-like agents increase negative responses, especially when the agent attempts to influence everyday actions.

The present study addresses these gaps by comparing dog, cat, and horse quadruped virtual agents in a mixed reality environment. Specifically, we compare species-specific behaviors, shared behaviors across species, and a bark-only baseline. This design allows us to examine whether persuasive effectiveness depends on faithful reproduction of animal-specific motion or on species-independent functional cues, such as emotional expression, attention guidance, and gaze alternation, that support presence and intention readability. Persuasive effectiveness is evaluated through multiple measures, including intention understanding, behavioral intention, actual behavior, psychological reactance, discomfort, familiarity, and acceptance.

Importantly, the comparison of multiple animal species is not merely a test of whether different agents are persuasive. Rather, species variation is used as a methodological tool to examine the mechanism of persuasive interaction. If persuasive effects are observed across different animal-like agents even when species-specific motions do not produce consistent advantages, this would suggest that persuasion is supported by more general functional cues rather than by animal-specific realism alone. In this sense, the present study uses cross-species comparison to clarify how emotional expression and attention guidance, together with presence and intention readability, contribute to behavior change.

By doing so, this study contributes to the development of Persuasive and Affective Human–AI Interaction in three ways. First, it extends research on pet-type persuasive agents from dog-like agents to multiple quadruped species. Second, it empirically examines whether persuasive behavior can be abstracted as species-independent functional design elements. Third, it provides design implications for standardizing persuasive behaviors in animal-like virtual agents, thereby supporting the development of reusable design patterns for non-humanoid affective persuasive interfaces.

## 3. MR-Based Experimental System and Persuasive Behavior Design

This section describes the experimental system and behavior design used to examine persuasive interaction with quadruped virtual agents in mixed reality. The purpose of the experiment was to investigate whether emotional expression and nonverbal persuasive movements can support intention understanding and behavior change across different animal-like agents, and whether species-specific motion enhances persuasive effectiveness compared with shared behavior patterns.

To address these questions, we developed an MR-based experimental system in which dog, cat, and horse virtual agents were presented in the participant's physical environment. The system allowed the experimenter to control the animal species, behavior condition, associated emotional expression, and start timing of each persuasive action. The agents performed persuasive behaviors related to everyday tasks, including trash disposal, feeding, and refraining from smartphone use.

The following subsections describe the experimental design in detail. First, we provide an overview of the experimental design and hypotheses. Next, we describe the MR system and experimental setup, the quadruped virtual agents, the design of persuasive behaviors, and the experimental conditions. We then explain the tasks and procedure, followed by the measures and data analysis methods.

### 3.1 Overview of the Experiment

The experiment was designed to examine how animal species and persuasive behavior design influence users' interpretation and behavioral responses to quadruped virtual agents in mixed reality. Specifically, the study addressed two research questions. First, can persuasive behaviors with emotional expression and nonverbal motion support intention understanding and behavior change not only in dog-like agents, but also in other quadruped animal-like agents? Second, does species-specific motion enhance persuasive effectiveness compared with shared behavior patterns that are applied across different animal species?

Based on these questions, we formulated two hypotheses. The first hypothesis was that persuasive behaviors with emotional expression would support intention understanding and behavior change across quadruped virtual agents, including non-dog agents. The second hypothesis was that species-specific behaviors would produce stronger persuasive effects than shared behaviors across species. These hypotheses correspond to the broader aim of examining whether persuasive behavior design should depend on animal-specific motion or whether it can be abstracted into species-independent functional design elements.

To test these hypotheses, we conducted a within-participants experiment using three quadruped virtual agents: a dog, a cat, and a horse. Each participant experienced persuasive interactions with the agents in an MR environment using Meta Quest 3. The agents attempted to influence participants in three everyday tasks: disposing of trash, feeding the agent, and refraining from smartphone use. These tasks were selected because they represent simple daily actions that can be influenced through attention guidance, emotional expression, and nonverbal communication.

The experiment included three behavior conditions. In the species-specific behavior condition, each animal agent performed persuasive movements designed to reflect its own species-specific characteristics. In the shared behavior condition, the agents performed common persuasive movements across species. In the bark-only condition, only the dog agent was presented and performed a barking animation without the full set of emotional expressions and structured persuasive movements. This condition was used as a minimal-expression baseline to examine whether the visual presence of a dog agent and simple barking alone were sufficient to communicate persuasive intention.

The main dependent measures included intention understanding, behavioral intention, actual behavior, psychological reactance, discomfort, familiarity, and agent acceptance. By comparing these measures across animal species and behavior conditions, the experiment aimed to clarify whether persuasive effectiveness depends on species-specific motion or on more general functional cues, such as emotional expression, attention guidance, and gaze alternation, that support intention readability.

**3.2 MR System and Experimental Setup**

We developed an MR-based experimental system to present quadruped virtual agents in the participant's physical environment. The system consisted of two components: an MR application experienced by the participant and a control application operated by the experimenter. These two applications were connected through UDP communication, allowing the experimenter to remotely control the animal species, behavior condition, start timing of persuasive actions, and reset of the agent's initial position, including the associated expressive behaviors.

The participant-side MR application was implemented in Unity and built for Meta Quest 3. We used the Meta XR All-in-One SDK to construct the mixed reality environment and to enable passthrough presentation. Through the passthrough function, participants could see the real room while the virtual animal agent was superimposed on the floor of the physical environment. This setup allowed the agent to perform persuasive behaviors in relation to real objects and locations, such as trash, a trash box, food, and a feeding dish.

The experimenter-side control application was implemented in Unity and operated on a laptop computer located outside the experimental room. From this control application, the experimenter sent string-based control signals to the MR application through UDP communication. When the MR application received a signal, it updated the internal state of the agent and selected the corresponding animation state. This enabled the experimenter to switch the animal species, select the behavior condition, trigger the persuasive action, and return the agent to its initial position without explicitly informing the participant of the experimental condition.

The virtual agents were displayed near the participant at the beginning of each persuasive interaction. The initial position was adjusted so that the agent appeared on the floor within the participant's field of view and could move toward relevant objects or locations during the task. The experimental room contained the objects required for the tasks, including a smartphone, trash, a trash box, food, and a feeding dish. After each condition, the experimenter returned the objects to their initial positions while the participant answered the questionnaire.

The experiment was conducted in a soundproof room at the university. Participants wore the Meta Quest 3 headset and experienced the persuasive behaviors of the virtual agents in the MR environment. Their behavior during the experiment was recorded using a smartphone camera and tripod for research purposes. The recorded data were anonymized and managed so that individual participants could not be identified.

This system allowed the virtual agents to be presented as situated entities in the participant's physical environment while maintaining experimental control over the timing, behavior condition, and animal species. Therefore, the system was suitable for examining how MR-based quadruped agents communicate persuasive intention through spatially grounded emotional and nonverbal behaviors. Figure 1 shows an example of the MR-based persuasive interaction in the trash disposal task. In this example, the dog agent is presented in the participant's physical environment and directs its body and attention toward the real trash box. This illustrates how the agent's persuasive behavior was grounded in real-world objects rather than being presented as a video or isolated animation.
The spatial arrangement of the participant, virtual agent, and task-related objects in the experimental room is shown in Supplementary Figure S1.

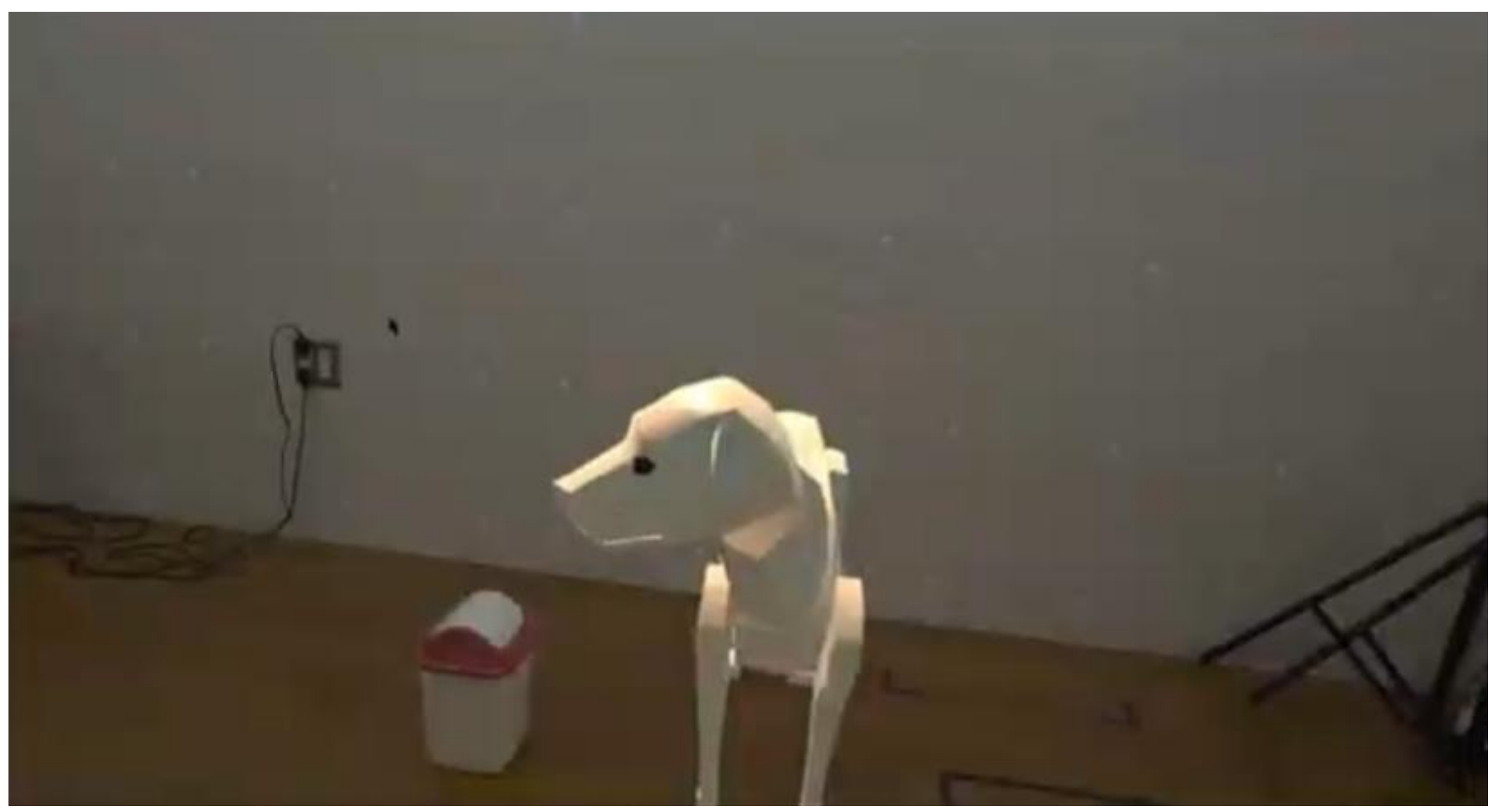

**Figure 1. Example of MR-based persuasive interaction in the trash disposal task.**

The dog agent was spatially presented in the participant's physical environment and directed its body and attention toward the real trash box. This screenshot illustrates how the agent's persuasive behavior was grounded in real-world objects rather than presented as a video or isolated animation.

### 3.3 Quadruped Virtual Agents

The experiment used three quadruped virtual agents representing a dog, a cat, and a horse. These animal species were selected because they share a quadruped body structure but differ in social familiarity, typical body scale, movement characteristics, and common human expectations. The dog was included as a familiar social animal and as a reference point for previous studies on dog-like persuasive agents. The cat and horse were included to examine whether persuasive behaviors designed for animal-like agents can generalize beyond dog-like communication. To reduce the possibility that perceived size would influence participants' interpretation or behavioral responses, the apparent display size of the dog, cat, and horse agents was adjusted to be comparable in the MR environment. Thus, body size was not intentionally manipulated as an experimental factor.

The 3D models and animations of the agents were created using Blender. For each animal species, we prepared a 3D model with a mesh and rig structure suitable for quadruped movement. The rigging process was supported by Rigify Zoo, a Blender add-on that provides animal-specific rig structures. The corresponding rig was applied to each animal model, and the mesh was parented to the rig so that the body, head, limbs, and other relevant parts could be animated.

Walking animations were based on existing animations provided by Rigify Zoo. In contrast, the animations related to persuasive behaviors were newly created using keyframe animation to match the purpose of the present study. These included behaviors for attracting attention, moving toward a target, looking at objects, alternating gaze or head orientation, and expressing emotional states. After the animations were created in Blender, the models, rigs, and animation data were exported in FBX format and imported into Unity.

In Unity, the animations were controlled using Animation Controllers. Each agent had animation states corresponding to waiting, moving, and persuasive actions. Transitions between these states were triggered by control signals sent from the experimenter-side application. Animation events were also used to synchronize vocal sounds with specific moments in the animation. This allowed the agents to present coordinated behaviors in which movement, orientation, emotional expression, and vocalization were temporally aligned.

Using this implementation, the dog, cat, and horse agents could be presented within the same MR application and controlled through a common interaction framework. This was important for the purpose of the study because it allowed the experiment to compare species-specific behaviors and shared behaviors while keeping the overall system, task environment, and control procedure consistent across animal species.

Figure 2 shows representative images of the dog, cat, and horse agents used in the experiment. The figure illustrates the 3D model of each quadruped agent and examples of representative expression states, including neutral or idle posture, an anger expression used as an attention-calling cue, and a sadness or need-related expression. In this study, the anger expression was used as a high-arousal cue to attract the participant's attention during the attention-calling phase, whereas the sadness or need-related expression was used to convey the agent's need or difficulty in relation to the target behavior. These images show how comparable affective and attentional cues were implemented across different animal species while preserving the same general persuasive functions.

| Agent species and 3D model | Neutral / idle | Attention-calling cue (anger expression) | Sadness / need-related expression |
|---|---|---|---|
| Dog | | | |
| Cat | | | |
| Horse | | | |

**Figure 2. Quadruped virtual agents and representative expression states used in the experiment. The figure shows the dog, cat, and horse virtual agents used in the experiment. The first column shows the 3D model of each agent, and the remaining columns show representative expression states, including neutral or idle posture, an anger expression used as an attention-calling cue, and a sadness or need-related expression. The anger expression was used only as an attention-calling cue, not as a general persuasive emotion throughout the interaction. These images illustrate how affective and attentional cues were implemented across different quadruped agents while preserving comparable persuasive functions. The screenshots are representative examples and do not exhaustively show all animations used in the experiment.**

### 3.4 Design of Persuasive Behaviors

The persuasive behaviors used in this study were designed as structured combinations of emotional expression and nonverbal movement. The purpose of these behaviors was not to give explicit verbal instructions, but to allow participants to infer the agent's intention from its affective and spatial behavior. Based on this design goal, each persuasive behavior was constructed from three functional components: attention calling, guiding, and pointing.

Attention calling was designed to attract the participant's attention and indicate that the agent was attempting to communicate. This component included behaviors such as looking toward the participant, vocalizing, approaching the participant, or turning back toward the participant during the sequence. These actions served as cues that the agent's behavior was directed toward the user rather than being random movement. In this study, an anger-like high-arousal expression was used only as part of the attention-calling phase, where the goal was to make the agent's signal salient rather than to express anger as the main persuasive emotion.

Guiding was designed to indicate the direction or location related to the target behavior. This component involved the agent moving toward a relevant place or object in the physical environment. For example, in the trash disposal task, the agent moved toward the location of the trash or the trash box. In the feeding task, the agent moved toward the feeding dish. Guiding behavior was intended to help participants understand where their attention or action should be directed.

Pointing was designed to indicate the object or relationship relevant to the target action. Because quadruped animal-like agents do not point with hands in the same way as humanoid agents, pointing was implemented through gaze direction, head orientation, and alternation between relevant objects. For example, the agent alternated its attention between the trash and the trash box, or between the food and the feeding dish. This gaze alternation was intended to help participants infer the relationship between objects and identify the action expected by the agent.

These three components were combined differently depending on the task. In the trash disposal task, the agent looked at the participant, vocalized, moved toward the trash, looked back at the participant, looked at the trash, and then looked at the trash box. In the feeding task, the agent similarly used attention calling, guiding, and pointing to indicate the relationship between the food and the feeding dish. In the smartphone-use task, the persuasive behavior was simpler and mainly consisted of looking at the participant and vocalizing, because the target behavior was not directly associated with moving an object to a visible destination.

Emotional expression was integrated into these persuasive behaviors to make the agent's internal state and need more interpretable. The expressions were designed to suggest affective states such as sadness, hunger, discomfort, or need, depending on the task. These expressions were not treated as decorative elements, but as functional cues that could help participants understand why the agent was acting and how they might respond. In this sense, the persuasive behaviors were designed to communicate not only a target action, but also an affective reason for action.

The design of these behaviors was informed by previous studies on pet-type persuasive agents. Harada and Sumi (2024) showed that emotional expressions such as sadness and confusion in a dog-like pet-type artifact could encourage users to perform a target behavior when the expressions were interpreted as signs of need or difficulty. This finding motivated the present study to treat emotional expression not as a decorative feature, but as a cue for communicating the agent's need, affective state, and persuasive intention. Sumi and Harada (2025) further examined persuasive interactions with pet-type virtual agents in mixed reality and suggested that the effectiveness of persuasion depends on the combination of emotional expression, context, and behavioral cues such as attention calling, guiding, and pointing.

Based on these findings, the present study adopted attention calling, guiding, and pointing as functional components of persuasive behavior and extended them to dog, cat, and horse agents. Thus, the behavior design was intended to separate the communicative function of a behavior from its animal-specific surface expression. This design allowed us to examine whether persuasive effectiveness depends on reproducing species-specific movements or on more general functional cues that support intention readability and affective interpretation.

Table 1 summarizes the structure of the persuasive behaviors used in each task. In the trash disposal and feeding tasks, the persuasive sequence included attention calling, guiding, and pointing. Specifically, the agent first attracted the participant's attention by looking at the participant and vocalizing, then moved toward a task-relevant object or location, and finally indicated the relationship between relevant objects through gaze direction or head orientation. In contrast, the smartphone-use task included only attention-calling behaviors, because the target behavior was not directly associated with moving an object to a visible destination.

**Table 1. Structure of persuasive behaviors used in each task.**

| Task | Step | Behavior | Functional component |
|---|---|---|---|
| Trash disposal | 1 | Looks at participant | Attention calling |
| Trash disposal | 2 | Vocalizes | Attention calling |
| Trash disposal | 3 | Moves toward trash | Guiding |
| Trash disposal | 4 | Looks back at participant | Attention calling |
| Trash disposal | 5 | Looks at trash | Pointing |
| Trash disposal | 6 | Looks at trash box | Pointing |
| Feeding | 1 | Looks at participant | Attention calling |
| Feeding | 2 | Vocalizes | Attention calling |
| Feeding | 3 | Moves toward feeding dish | Guiding |
| Feeding | 4 | Alternates gaze between food and dish | Pointing |
| Smartphone use | 1 | Looks at participant | Attention calling |
| Smartphone use | 2 | Vocalizes | Attention calling |

This difference in behavioral structure reflects the spatial characteristics of each task: trash disposal and feeding could be grounded in visible objects and destinations, whereas refraining from smartphone use required the participant to infer the agent's intention mainly from attention-calling and affective cues.

### 3.5 Experimental Conditions

The experiment included three behavior conditions: the species-specific behavior condition, the shared behavior condition, and the bark-only condition. These conditions were designed to examine whether persuasive effectiveness depends on species-specific motion or on more general functional cues, such as emotional expression, attention guidance, and pointing, that support intention readability. Table 2 summarizes the agent species, condition labels used in the results, main behavior, and purpose of each behavior condition.

**Table 2. Experimental behavior conditions.**

| Condition | Agent species | Condition labels used in the results | Main behavior | Purpose |
|---|---|---|---|---|
| Species-specific behavior | Dog, cat, horse | U-dog, U-cat, U-horse | Persuasive behaviors using attention calling, guiding, and pointing, with surface motions adapted to each species' expressive style | To examine whether species-specific expressive motion enhances persuasion |
| Shared behavior | Dog, cat, horse | C-dog, C-cat, C-horse | Persuasive behaviors using the same functional sequence of attention calling, guiding, and pointing across species | To examine whether persuasive behaviors can be standardized as functional cues |
| Bark-only | Dog only | B-dog | Visible dog agent performing only a barking animation | Minimal-expression baseline |

As shown in Table 2, the species-specific and shared behavior conditions were implemented for all three animal species, whereas the bark-only condition was implemented only for the dog agent. The condition labels shown in the table are used in the results section: U refers to unique, species-specific behavior, C refers to shared behavior, and B refers to the bark-only baseline. The species-specific and shared behavior conditions both included structured persuasive behaviors composed of attention calling, guiding, and pointing. However, they differed in whether the surface form of these functions was adapted to each species' expressive style or kept as a common persuasive sequence across species. The bark-only condition was implemented only for the dog agent because it was designed as a baseline derived from previous dog-like agent studies, and because barking is a familiar communicative cue in human–dog interaction.

In the species-specific behavior condition, each animal agent performed persuasive behaviors that were designed to reflect characteristics associated with its own species. The dog, cat, and horse agents used movements and expressive styles intended to appear appropriate to their respective animal forms. In this condition, the communicative functions of attention calling, guiding, and pointing were preserved across species, but their surface motions were adapted to each animal's characteristic movement style. The dog agent used relatively direct approach behavior, clear orientation toward the participant or target object, and dog-like vocalization. The cat agent used more cautious and indirect approach behavior, subtler body orientation, and smaller head or body movements to convey attention and need. The horse agent used broader body movements and more salient head orientation to indicate direction and attention, reflecting the larger movement impression commonly associated with horses. These behaviors were not intended to reproduce biologically accurate animal behavior in detail. Rather, they were designed to provide species-appropriate expressive variations of the same functional persuasive components. This condition was included to examine whether species-specific expressive motion enhances persuasive effectiveness.

In the shared behavior condition, the dog, cat, and horse agents performed persuasive behaviors based on a common behavioral structure. The same functional components, such as attention calling, guiding, and pointing, were applied across animal species. In contrast to the species-specific behavior condition, the timing and sequence of attention calling, guiding, and pointing were kept as consistent as possible across the dog, cat, and horse agents, so that the same persuasive structure was applied regardless of animal species. The surface motion was adjusted to each animal's body structure when necessary, but the communicative function and sequence of the behavior were maintained. This condition was included to examine whether persuasive behaviors can be standardized as species-independent functional cues.

In the bark-only condition, only the dog agent was used. The dog agent was visually presented in the MR environment, but it performed only a barking animation without the full set of emotional expressions, gaze alternation, guiding movements, pointing behaviors, or structured persuasive sequences used in the other conditions. This condition served as a minimal-expression baseline. It was designed to examine whether the visual presence of a dog agent and simple barking alone were sufficient to communicate persuasive intention and promote behavior.

The bark-only condition should therefore not be understood as an audio-only condition. Participants could see the dog agent, but the agent did not perform the structured affective and nonverbal behaviors used in the species-specific and shared behavior conditions. This distinction is important because the comparison with the bark-only condition allowed us to evaluate the contribution of emotional expression and structured persuasive movement beyond the mere presence of an animal-like agent.

The full set of experimental conditions consisted of seven combinations: dog with species-specific behavior, dog with shared behavior, dog with bark-only behavior, cat with species-specific behavior, cat with shared behavior, horse with species-specific behavior, and horse with shared behavior. These seven combinations were experienced by each participant in a within-participants design. The bark-only condition was implemented only for the dog agent because it was designed as a baseline derived from previous dog-like agent studies and because barking is a familiar communicative cue for dogs in human–dog interaction.

### 3.6 Tasks and Procedure

The experiment used three everyday tasks as persuasive scenarios: trash disposal, feeding, and refraining from smartphone use. In this study, spatial grounding refers to the extent to which the target behavior can be indicated through visible physical objects, target locations, and spatial relationships in the experimental room. These tasks were selected because they differ in the degree to which the target action can be spatially grounded in the physical environment. In the trash disposal task, the participant was expected to infer that the agent wanted them to throw away the trash. In the feeding task, the participant was expected to infer that the agent wanted to be fed. In the smartphone-use task, the participant was expected to infer that the agent wanted them to stop using or refrain from using the smartphone. Table 3 summarizes the target behavior, spatial grounding, and main objects involved in each task.

**Table 3. Experimental tasks and spatial grounding.**

| Task | Target behavior | Spatial grounding | Main objects |
|---|---|---|---|
| Trash disposal | Throw away trash | High | Trash, trash box |
| Feeding | Feed the agent | High | Food, feeding dish |
| Refraining from smartphone use | Stop or refrain from smartphone use | Low | Smartphone |

As shown in Table 3, the trash disposal and feeding tasks involved visible objects and target locations, whereas the smartphone-use task was less directly associated with moving an object to a visible destination. This distinction was important because the persuasive behaviors in this study relied on spatially grounded cues, such as gaze direction, guiding, and pointing, in addition to emotional expression.

The experiment was conducted with 16 undergraduate and graduate students at a university. A within-participants design was used, meaning that each participant experienced all seven combinations of animal species and behavior condition described in Section 3.5. Before the experiment, participants received an explanation of the study and provided informed consent. Each participant was assigned an anonymous participant ID, and all data were managed so that individuals could not be identified. Participants received compensation according to the university's regulations.

At the beginning of the experiment, participants wore the Meta Quest 3 headset and experienced the MR application in the experimental room. They were informed that they would experience behavior-support agents in mixed reality. To help participants distinguish the agents and to make the interaction more natural, each virtual agent was introduced with a name: the dog was named Taro, the cat was named Tama, and the horse was named Cookie. The names were not intended as an experimental manipulation, but as a way to present each agent as an identifiable interaction partner. Participants were instructed to remain in place until the persuasive behavior began. After the agent completed its persuasive behavior, participants were allowed to act freely, including touching or moving the objects placed in the room.

After each persuasive interaction, participants answered a questionnaire using Google Forms. The behavior conditions were recorded using labels such as U, C, and B for data management, corresponding to species-specific behavior, shared behavior, and bark-only behavior, respectively. The meanings of these labels were not explained to the participants.

The presentation order was partially controlled to reduce order effects. The order of animal species and behavior conditions was counterbalanced across participants based on a prearranged assignment table. The smartphone-use task was always presented first because it required participants to have the smartphone available at the beginning of the interaction and because later interaction with the physical objects used in the trash disposal and feeding tasks could have affected the setup. The order of the remaining two tasks, trash disposal and feeding, was counterbalanced across participants.

After each condition, while the participant completed the questionnaire, the experimenter returned the physical objects, such as trash, the trash box, food, and the feeding dish, to their initial positions. If the headset position shifted when the participant adjusted or re-wore the Meta Quest 3, the experimenter corrected the initial position of the virtual agent before the next trial. The entire experiment took approximately one hour, including the explanation, MR experience, questionnaires, and debriefing. This procedure enabled comparison of participants' intention understanding, behavioral intention, actual behavior, and subjective evaluations across animal species and behavior conditions in a controlled MR environment.

### 3.7 Measures

After each persuasive interaction, participants completed a questionnaire using Google Forms. The questionnaire was designed to evaluate how participants interpreted the agent's behavior, whether they intended to perform the target action, whether they actually performed the action, and how they subjectively evaluated the persuasive interaction. The same set of questions was used for each animal agent and each task, with the wording adjusted to match the animal species and target behavior.

The first measure was familiarity with the animal. Participants rated how familiar they felt with the animal species on a 7-point Likert scale. This measure was included to examine whether participants' prior affinity toward each animal species was related to their behavioral responses to the agent.

The second measure was intention understanding. For each task, participants answered whether they felt that the agent wanted them to perform the target behavior, such as throwing away trash, feeding the agent, or refraining from smartphone use. This item was answered using a yes/no format. Intention understanding was used as a central measure because the present study focused on whether emotional expression and nonverbal behavior made the agent's persuasive intention readable.

The third and fourth measures were psychological reactance and discomfort. Participants rated the extent to which they felt resistance or discomfort toward the agent's attempt to influence their behavior on 7-point Likert scales. These measures were included to examine whether persuasive behaviors by quadruped virtual agents elicited negative psychological responses or increased psychological burden. The fifth measure was perceived emotion of the agent. Participants selected what kind of emotion they thought the agent expressed during each persuasive behavior from multiple-choice options and could also provide an open-ended response. This measure was used to examine how participants interpreted the agent's affective state and whether the emotional expression was understood in relation to the target behavior.

The sixth measure was behavioral intention. For each task, participants answered whether they wanted to perform the target behavior after observing the agent's persuasive behavior. This item was answered using a yes/no format. Behavioral intention was included to assess whether the agent's behavior influenced participants' willingness to act.
The seventh measure was actual behavior. For each task, participants answered whether they actually performed the target behavior during the experiment. In addition to questionnaire responses, participants' behavior was recorded using a smartphone camera for research purposes. Actual behavior was used as a behavioral outcome measure to examine whether the agent's persuasive behavior led to action, not only to intention.

The eighth measure was agent acceptance. Participants rated the extent to which they wanted to stay with the agent on a 7-point Likert scale. This item was used as an indicator of acceptance or affinity toward the agent after the interaction.

The questionnaire also included open-ended questions. When participants reported that they performed the target behavior, they were asked to describe why they did so. Participants were also asked how the agent's motion or emotional expression could be improved to make them more likely to perform the target behavior. These free descriptions were used to interpret the quantitative results and to identify design implications, such as the roles of emotional expression, gaze alternation, attention guidance, and post-action feedback.

Table 4 summarizes the questionnaire items used in the experiment. The same item structure was used across animal species and tasks, with the animal name and target behavior replaced according to the condition. For example, the intention-understanding item asked whether the agent seemed to want the participant to perform the target behavior, such as "Did the dog seem to want you to throw away the trash?" Similarly, behavioral intention and actual behavior were measured by asking whether the participant wanted to perform the target behavior and whether they actually performed it.

**Table 4. Questionnaire items used in the experiment.**

| Measure | Example item | Response format |
|---|---|---|
| Familiarity | How familiar do you feel with dogs/cats/horses? | 7-point Likert scale |
| Intention understanding | Did the agent seem to want you to [throw away the trash/feed it/refrain from smartphone use]? | Yes/No |
| Psychological reactance | To what extent did you feel resistance toward the agent's attempt to make you [target behavior]? | 7-point Likert scale |
| Discomfort | To what extent did you feel discomfort toward the agent's attempt to make you [target behavior]? | 7-point Likert scale |
| Perceived emotion | What emotion did the agent seem to express? | Multiple choice / open-ended |
| Behavioral intention | Did you want to [target behavior]? | Yes/No |
| Actual behavior | Did you actually [target behavior]? | Yes/No |
| Reason for action | Why did you perform the behavior? | Open-ended |
| Improvement | How could the agent's motion or emotional expression be improved to make you want to perform the behavior? | Open-ended |
| Agent acceptance | To what extent did you want to stay with the agent? | 7-point Likert scale |

### 3.8 Data Analysis

The data analysis was conducted to examine the effects of animal species and behavior condition on participants' interpretation, behavioral responses, and subjective evaluations. The experimental conditions consisted of seven combinations: dog with species-specific behavior, dog with shared behavior, dog with bark-only behavior, cat with species-specific behavior, cat with shared behavior, horse with species-specific behavior, and horse with shared behavior. For clarity in the results, these conditions were abbreviated as U-dog, C-dog, B-dog, U-cat, C-cat, U-horse, and C-horse, respectively.

For binary measures, including intention understanding, behavioral intention, and actual behavior, the proportion of "yes" responses was calculated for each condition and task. Condition differences were examined separately for each task using Fisher's exact tests. When an overall difference among conditions was found, multiple comparisons were conducted to identify which conditions differed from each other. These analyses were used to examine whether the bark-only condition differed from the conditions involving emotional expression and structured persuasive behavior, and whether species-specific behavior differed from shared behavior within each animal species.

For 7-point rating measures, including familiarity, psychological reactance, discomfort, and agent acceptance, descriptive statistics were calculated for each condition. Condition differences were examined using repeated-measures analyses of variance. When significant condition effects were found, post hoc comparisons were conducted to examine differences among specific conditions. Psychological reactance and discomfort were analyzed to evaluate whether persuasive behaviors increased negative psychological responses, whereas familiarity and agent acceptance were analyzed to examine users' subjective relationship to each agent.

To explore the relationship between familiarity and actual behavior, logistic regression analyses were conducted. Actual behavior was treated as a binary dependent variable, with 0 indicating that the participant did not perform the target behavior and 1 indicating that the participant performed the target behavior. Familiarity ratings were used as the independent variable. These analyses were conducted separately for each animal species and behavior-condition group to examine whether familiarity predicted the likelihood of actual behavior.

Free descriptions were reviewed qualitatively to support the interpretation of the quantitative results and to extract design implications. Responses to the question asking why participants performed the target behavior were reviewed and grouped into categories such as emotional expression, gaze or attention guidance, vocalization, perceived need of the agent, and task-related understanding. Responses to the question asking how the agent's behavior could be improved were reviewed to identify suggestions such as clearer intention expression, stronger attention guidance, closer spatial approach, and post-action feedback. These qualitative results were used to interpret how participants understood the agents' intentions and what design elements may improve persuasive effectiveness.

Statistical significance was evaluated using a conventional threshold of $p < .05$. When multiple comparisons were conducted, p-values were adjusted using the Holm method to reduce the risk of Type I error. The quantitative and qualitative analyses were interpreted together to examine not only whether persuasive effects occurred, but also how participants understood the agents' emotional expressions, nonverbal behaviors, and persuasive intentions.

**4. Results**

This section presents the results of the experiment on persuasive behaviors by quadruped virtual agents. The analyses examined how animal species and behavior condition influenced participants' intention understanding, behavioral intention, actual behavior, psychological reactance, discomfort, familiarity, and agent acceptance across the three tasks: trash disposal, feeding, and refraining from smartphone use. The following subsections summarize the main findings and then report the detailed results for each measure.

### 4.1 Overview of the Results

The analysis focused on seven combinations of animal species and behavior condition: dog with species-specific behavior, dog with shared behavior, dog with bark-only behavior, cat with species-specific behavior, cat with shared behavior, horse with species-specific behavior, and horse with shared behavior. For clarity, these conditions are abbreviated as U-dog, C-dog, B-dog, U-cat, C-cat, U-horse, and C-horse, respectively.

The results are organized as follows. First, Section 4.2 reports the effects of expressive persuasive behaviors on intention understanding, behavioral intention, and actual behavior, with particular focus on comparisons with the bark-only baseline. These analyses address Hypothesis 1. Figure 3 visualizes these behavioral outcome measures across conditions and tasks. Second, Section 4.3 compares species-specific behaviors with shared behaviors and reports familiarity ratings across conditions. These analyses address Hypothesis 2. Third, Section 4.4 reports psychological reactance, discomfort, and agent acceptance. Section 4.5 examines the relationship between familiarity and actual behavior using logistic regression. Finally, Section 4.6 reports the qualitative findings from free-description responses.

Table 5 summarizes the main results. Overall, the bark-only baseline produced lower intention understanding in the trash disposal and feeding tasks, and lower behavioral intention and actual behavior particularly in the feeding task. In contrast, no consistent significant differences were found between species-specific and shared behavior conditions. Psychological reactance and discomfort remained low across conditions, and agent acceptance did not differ significantly among conditions. Exploratory logistic regression analyses showed that familiarity positively predicted actual behavior in the cat and horse conditions and in the dog bark-only condition.

**Table 5. Summary of the main results.**

| Measure | Main statistical result | Summary of result |
|---|---|---|
| Intention understanding | Significant differences among conditions were found in the trash disposal and feeding tasks. The bark-only condition was significantly lower than the other conditions in Holm-adjusted multiple comparisons. No significant difference among conditions was found in the smartphone-use task. | The bark-only baseline showed lower intention understanding, especially in spatially grounded tasks. |
| Behavioral intention | A significant difference among conditions was found in the trash disposal task, but Holm-adjusted multiple comparisons did not identify specific significant pairs. In the feeding task, the bark-only condition was significantly lower than the dog species-specific and dog shared behavior conditions. No significant difference among conditions was found in the smartphone-use task. | Behavioral intention tended to be lower in the bark-only condition, particularly in the feeding task. |
| Actual behavior | A significant difference among conditions was found in the trash disposal task, but Holm-adjusted multiple comparisons did not identify specific significant pairs. In the feeding task, the bark-only condition was significantly lower than all other conditions. No significant difference among conditions was found in the smartphone-use task. | Actual behavior was lower in the bark-only condition, especially in the feeding task. |
| Species-specific vs. shared behavior | No consistent significant differences were found between species-specific and shared behavior conditions across the main behavioral and subjective measures. | Species-specific behavior did not consistently outperform shared behavior. |
| Familiarity | Familiarity differed significantly across conditions, but no significant differences were found between species-specific and shared behavior conditions within the same animal species. | Familiarity differences mainly reflected animal species rather than behavior type. |
| Psychological reactance | No significant condition effects were found in any task. | Reactance remained low across conditions. |
| Discomfort | No significant condition effects were found in any task. | Discomfort remained low across conditions. |

| Agent acceptance | No significant condition effect was found. | Agent acceptance did not differ significantly across conditions. |
|---|---|---|
| Familiarity and actual behavior | Exploratory logistic regression analyses showed that familiarity positively predicted actual behavior in the cat and horse conditions and in the dog bark-only condition. | Familiarity was associated with actual behavior in several conditions. |
| Free descriptions | Participants frequently referred to emotional expression, gaze and motion cues, vocalization, difficulty in understanding intention, and the need for post-action feedback. | Free descriptions highlighted emotional expression, attention guidance, and feedback as relevant design elements. |

### 4.2 Intention Understanding, Behavioral Intention, and Actual Behavior

We first examined whether expressive and structured persuasive behaviors improved participants' intention understanding, behavioral intention, and actual behavior compared with the bark-only baseline. This analysis corresponds to Hypothesis 1, which predicted that persuasive behaviors with emotional expression would support intention understanding and behavior change across quadruped virtual agents.

Figure 3 visualizes the proportions of intention understanding, behavioral intention, and actual behavior across conditions and tasks. Overall, the bark-only baseline showed lower values than the expressive behavior conditions in several measures, particularly for the trash disposal and feeding tasks. In contrast, the smartphone-use task showed less consistent differences across conditions.

For intention understanding, significant differences among conditions were found in the trash disposal and feeding tasks. In the trash disposal task, intention understanding was 18.8% in the bark-only condition, whereas the expressive behavior conditions ranged from 87.5% to 100.0%. In the feeding task, intention understanding was also 18.8% in the bark-only condition, whereas the expressive behavior conditions ranged from 56.3% to 100.0%. Holm-adjusted multiple comparisons showed that the bark-only condition was significantly lower than all other conditions in both tasks. No significant difference among conditions was found in the smartphone-use task.

For behavioral intention, a significant difference among conditions was found in the trash disposal task, although Holm-adjusted multiple comparisons did not identify specific significant differences between pairs of conditions. In the feeding task, a significant difference among conditions was also found. Holm-adjusted multiple comparisons showed that the bark-only condition was significantly lower than the dog species-specific and dog shared behavior conditions. No significant difference among conditions was found in the smartphone-use task.

For actual behavior, a significant difference among conditions was found in the trash disposal task, although Holm-adjusted multiple comparisons did not identify specific significant differences between pairs of conditions. In the feeding task, a significant difference among conditions was found, and Holm-adjusted multiple comparisons showed that the bark-only condition was significantly lower than all other conditions. In this task, no participant performed the target behavior in the bark-only condition, whereas actual behavior rates in the expressive behavior conditions ranged from 50.0% to 62.5%. No significant difference among conditions was found in the smartphone-use task.

Taken together, these results show that the bark-only baseline was less effective than the expressive persuasive behavior conditions, particularly for intention understanding and for behavioral outcomes in the feeding task. Therefore, Hypothesis 1 was partially supported. Emotional expression and structured nonverbal behaviors appeared to support intention understanding and behavior change in several tasks, but their effects were not uniform across all task contexts.

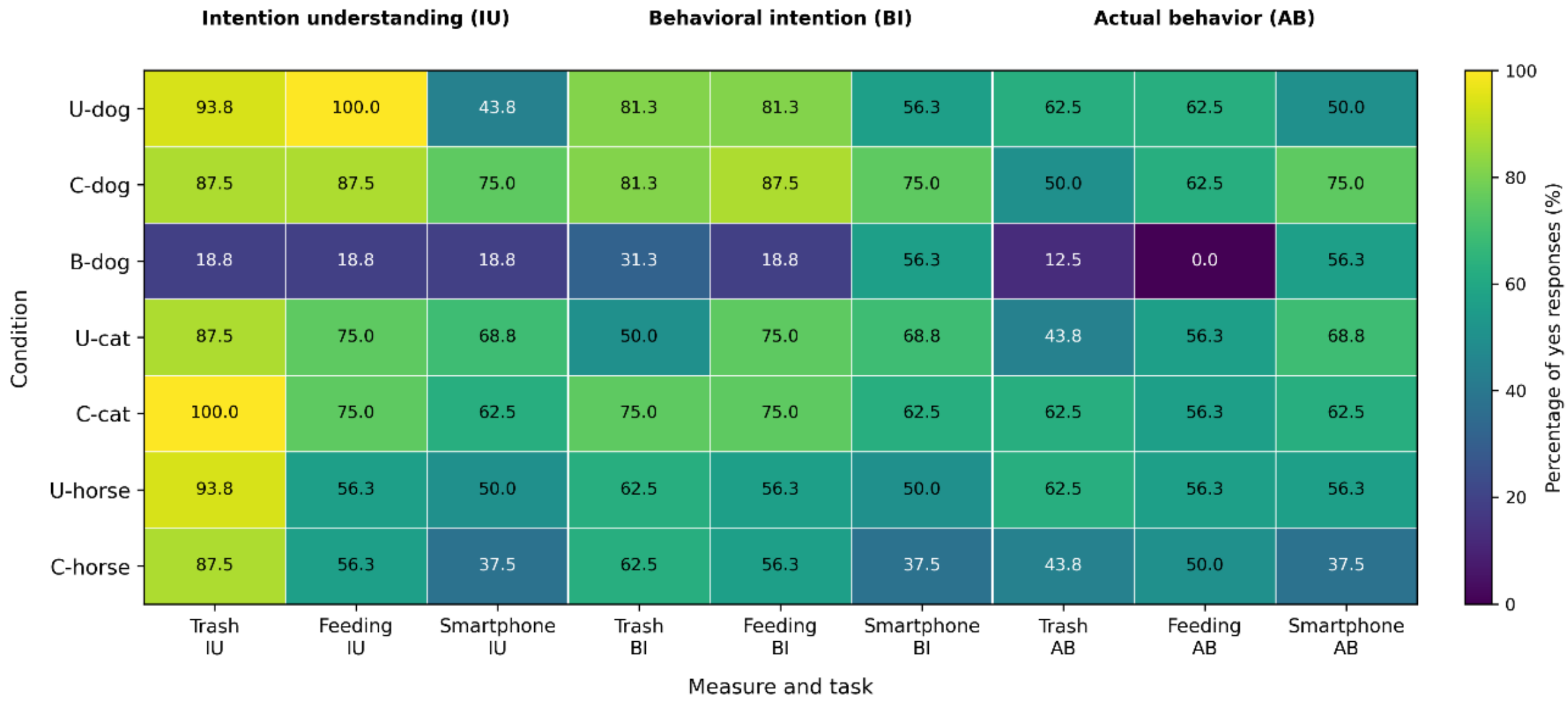

**Figure 3. Heatmap of behavioral outcome measures across conditions and tasks.**

The heatmap shows the proportions of "yes" responses for intention understanding, behavioral intention, and actual behavior across the seven experimental conditions and three tasks. Darker cells indicate higher proportions. IU = intention understanding; BI = behavioral intention; AB = actual behavior. U = species-specific behavior; C = shared behavior; B = bark-only baseline. Dog, cat, and horse indicate the agent species.

**4.3 Species-Specific Behaviors Versus Shared Behaviors**

We next examined whether species-specific behaviors produced stronger persuasive effects than shared behaviors. This analysis corresponds to Hypothesis 2, which predicted that persuasive behaviors designed to reflect each animal species would be more effective than common persuasive behaviors applied across species. In this section, the bark-only condition was excluded from the main comparison of species-specific and shared behaviors, except when reporting the overall descriptive statistics for familiarity.

Figure 4 shows the mean familiarity ratings for each condition. Familiarity differed significantly across the seven conditions, $F(6, 90) = 9.25$, $p < .001$. Post hoc comparisons showed that C-cat was rated significantly higher than C-horse, $p = .037$, and that U-dog was rated significantly higher than C-horse, $p = .037$. However, no significant differences were found between the species-specific and shared behavior conditions within the same animal species. These results indicate that the differences in familiarity were mainly related to animal species rather than to whether the behavior was species-specific or shared.

For intention understanding, both the species-specific and shared behavior conditions showed relatively high scores in the trash disposal and feeding tasks, as shown in Figure 3. However, species-specific behaviors did not consistently produce higher intention understanding than shared behaviors. In the trash disposal task, intention understanding was 93.8% for U-dog and 87.5% for C-dog, 87.5% for U-cat and 100.0% for C-cat, and 93.8% for U-horse and 87.5% for C-horse. In the feeding task, intention understanding was 100.0% for U-dog and 87.5% for C-dog, 75.0% for both U-cat and C-cat, and 56.3% for both U-horse and C-horse. In the smartphone-use task, no significant condition effect was found. Thus, species-specific motion did not provide a consistent advantage in communicating the agents' persuasive intentions.

A similar pattern was observed for behavioral intention and actual behavior. Although significant overall condition effects were found for some task measures, post hoc comparisons did not identify consistent significant differences between species-specific and shared behavior conditions. For behavioral intention, the trash disposal task showed a significant overall condition effect, but no specific pairwise differences were found. In the feeding task, the bark-only condition was significantly lower than the dog species-specific and dog shared behavior conditions, but no significant difference was found between U-dog and C-dog. For actual behavior, the trash disposal task also showed a significant overall condition effect, but no specific pairwise differences were found. In the feeding task, the bark-only condition was significantly lower than all other conditions, whereas no consistent differences were observed between species-specific and shared behavior conditions.

Taken together, these results do not support Hypothesis 2. In the present experiment, species-specific behaviors did not consistently enhance persuasive effectiveness compared with shared behaviors. Although familiarity differed among conditions, the significant differences were found mainly between animal species, especially between dog or cat conditions and the horse shared behavior condition, rather than between species-specific and shared behaviors within the same animal species.

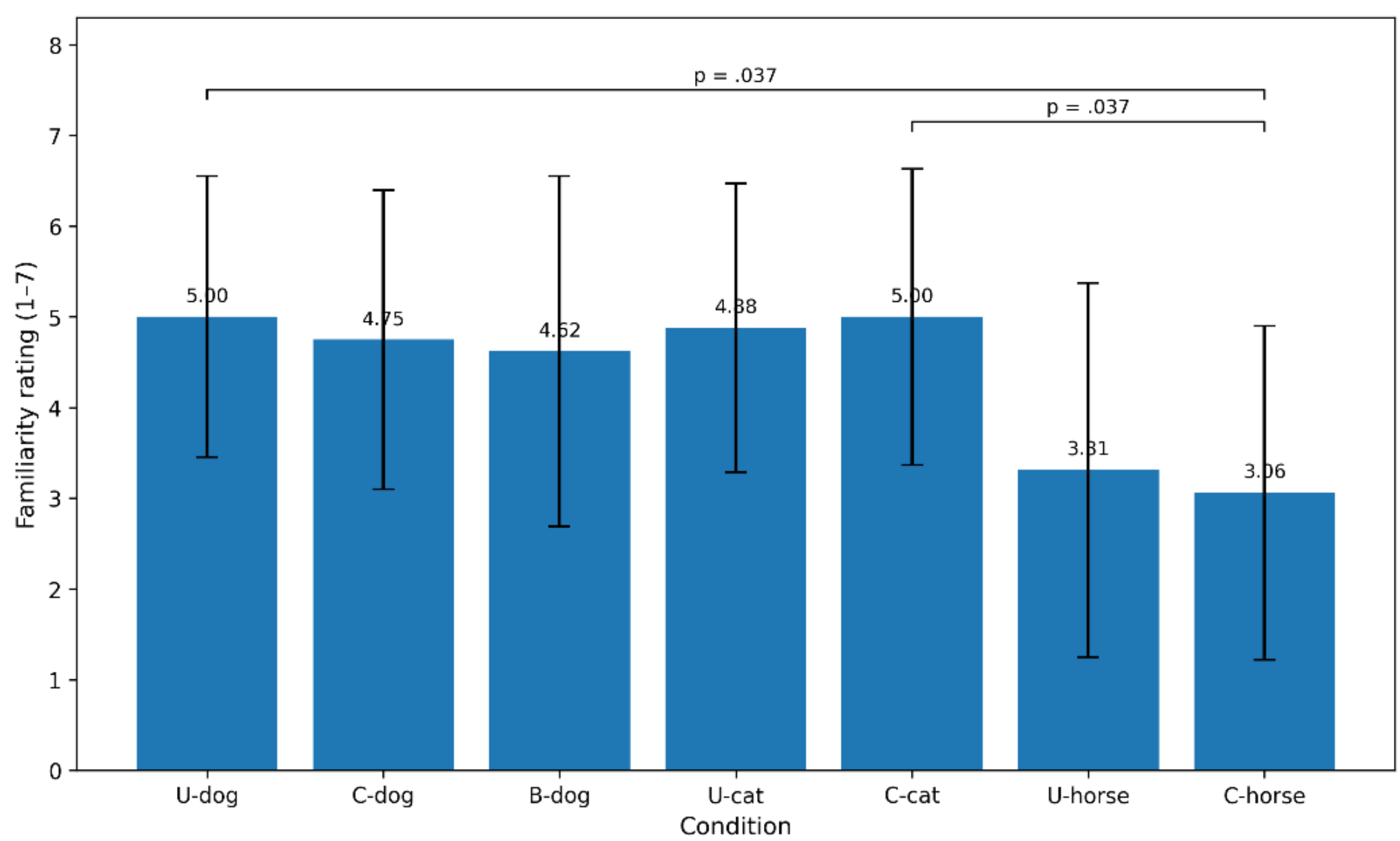

**Figure 4. Familiarity ratings across conditions.**

Bars indicate mean familiarity ratings on a 7-point scale, and error bars indicate standard deviations. Familiarity differed significantly across conditions, $F(6, 90) = 9.25$, $p < .001$. Post hoc comparisons showed that C-cat and U-dog were rated significantly higher than C-horse, $p = .037$. U = species-specific behavior; C = shared behavior; B = bark-only baseline. Dog, cat, and horse indicate the agent species.

### 4.4 Psychological Reactance, Discomfort, and Acceptance

We then examined whether the persuasive behaviors elicited negative psychological responses or influenced participants' acceptance of the agents. Psychological reactance and discomfort were measured for each of the three tasks using 7-point rating scales. Agent acceptance was measured by asking participants how much they wanted to stay with the agent.

For psychological reactance, no significant condition effects were found in any of the three tasks: trash disposal, $F(6, 90) = 0.45$, $p = .846$; feeding, $F(6, 90) = 0.39$, $p = .883$; and smartphone use, $F(6, 90) = 1.12$, $p = .356$. Across all tasks and conditions, the mean reactance ratings ranged from 1.81 to 2.88 on the 7-point scale. These results indicate that the persuasive behaviors did not elicit strong resistance from participants.

For discomfort, no significant condition effects were found in any of the three tasks: trash disposal, $F(6, 90) = 0.96$, $p = .454$; feeding, $F(6, 90) = 1.09$, $p = .375$; and smartphone use, $F(6, 90) = 1.55$, $p = .171$. Across all tasks and conditions, the mean discomfort ratings ranged from 1.44 to 2.38. Thus, neither the expressive persuasive behavior conditions nor the bark-only baseline produced high levels of discomfort.

Agent acceptance also did not differ significantly across conditions, $F(6, 90) = 1.78$, $p = .112$. The mean ratings for wanting to stay with the agent ranged from 3.38 to 4.69. Although the dog conditions tended to show relatively higher acceptance ratings than the horse conditions, the differences were not statistically significant.
Taken together, these results show that the persuasive behaviors used in this study did not increase psychological reactance or discomfort, and that agent acceptance remained moderate across conditions. Therefore, the differences observed in intention understanding and behavioral outcomes cannot be explained by strong negative psychological responses toward specific agents or behavior conditions.

### 4.5 Familiarity and Actual Behavior

We further examined whether participants' familiarity with each animal was related to actual behavior after the persuasive interaction. This analysis was exploratory and was conducted to examine whether prior affinity toward an animal species influenced the likelihood of performing the target behavior.
Logistic regression analyses were conducted separately for each animal species and behavior-condition group. Actual behavior was treated as the dependent variable, with 0 indicating that the participant did not perform the target behavior and 1 indicating that the participant performed the target behavior. Familiarity rating was used as the independent variable.

Table 6 summarizes the logistic regression results. Familiarity did not significantly predict actual behavior in the dog conditions with structured persuasive behaviors, OR = 1.10, $p = .482$. In contrast, familiarity significantly and positively predicted actual behavior in the cat conditions with structured persuasive behaviors, OR = 1.66, $p = .001$. Similarly, familiarity significantly and positively predicted actual behavior in the horse conditions with structured persuasive behaviors, OR = 1.91, $p < .001$. In the dog bark-only condition, familiarity also significantly and positively predicted actual behavior, OR = 1.64, $p = .038$.

These results indicate that participants who felt more familiar with cats or horses were more likely to perform the target behaviors after interacting with those agents. Familiarity was also associated with actual behavior in the dog bark-only condition, where persuasive cues were limited. In contrast, familiarity did not significantly predict actual behavior for the dog agent when structured persuasive behaviors were present.

**Table 6. Logistic regression results predicting actual behavior from familiarity.**

| Condition group | Odds ratio | p-value | Result |
|---|---|---|---|
| Dog with structured persuasive behaviors | 1.10 | .482 | Not significant |
| Cat with structured persuasive behaviors | 1.66 | .001 | Significant positive effect |
| Horse with structured persuasive behaviors | 1.91 | < .001 | Significant positive effect |
| Dog bark-only baseline | 1.64 | .038 | Significant positive effect |

### 4.6 Qualitative Findings

We reviewed participants' free-description responses to further understand why they performed the target behaviors and how the agents' persuasive behaviors could be improved. The free-description questions asked participants to explain why they performed the target behavior and how the agent's motion or emotional expression could be improved to make them more likely to perform the behavior. The responses were reviewed and grouped into recurring themes, including emotion and empathy, gaze and motion cues, intention inference, auditory or vocal cues, feedback, clearer motion, and physical or spatial intervention.

Table 7 shows the categorized responses regarding the reasons for performing the target behavior. Across the three tasks, emotion- and empathy-related responses were frequently observed. In the trash disposal task, participants often mentioned that the agent looked sad or uncomfortable. In the feeding task, many participants stated that the agent appeared hungry, sad, or in need of help. In the smartphone-use task, participants also referred to the agent appearing lonely or wanting attention. These responses indicate that participants interpreted the agents' emotional states and used such interpretations as a basis for action.

Gaze and motion cues were also frequently mentioned. In the trash disposal task, participants referred to the agent looking alternately at the trash and the trash box or directing attention toward the trash. In the feeding task, participants mentioned that the agent looked at the food or feeding dish. These responses indicate that gaze alternation and object-directed motion helped participants infer the relationship between the target object and the expected action. In the smartphone-use task, some participants referred to vocalization or the agent's apparent desire for attention, suggesting that social and affective interpretations were also involved when the task was less spatially grounded.

**Table 7. Categorized free-description responses for reasons for action.**

Values indicate the number of open-ended responses assigned to each category. For example, a value of 5 in the U-dog column for Emotion/empathy in the trash disposal task means that five responses in that condition referred to emotional or empathic reasons for performing the target behavior.

| Task | Category | U-dog | C-dog | B-dog | U-cat | C-cat | U-horse | C-horse | Total |
|---|---|---|---|---|---|---|---|---|---|
| Trash disposal | Emotion/empathy | 5 | 4 | 1 | 3 | 2 | 2 | 3 | 20 |
| Trash disposal | Gaze/motion | 4 | 1 | 0 | 2 | 0 | 2 | 1 | 10 |
| Trash disposal | Intention inference | 1 | 0 | 0 | 0 | 0 | 1 | 2 | 4 |
| Trash disposal | Other | 0 | 2 | 0 | 2 | 6 | 4 | 1 | 15 |
| Feeding | Emotion/empathy | 6 | 6 | 0 | 6 | 1 | 2 | 1 | 22 |
| Feeding | Gaze/motion | 3 | 1 | 0 | 1 | 1 | 4 | 1 | 11 |
| Feeding | Intention inference | 0 | 1 | 0 | 1 | 2 | 1 | 2 | 7 |
| Feeding | Other | 1 | 2 | 0 | 1 | 3 | 2 | 4 | 13 |
| Smartphone use | Emotion/empathy | 0 | 4 | 3 | 4 | 4 | 1 | 0 | 16 |
| Smartphone use | Gaze/motion | 0 | 5 | 1 | 2 | 1 | 1 | 4 | 14 |
| Smartphone use | Auditory/vocal cues | 0 | 1 | 1 | 1 | 1 | 1 | 0 | 5 |
| Smartphone use | Other | 5 | 2 | 3 | 3 | 3 | 5 | 2 | 23 |

Table 8 shows the categorized responses regarding suggested improvements. Across tasks, participants frequently requested clearer feedback after they performed the target behavior. For example, several responses suggested that the agent should react after being fed, show that it had accepted the user's action, or provide a visible emotional response after the participant completed the requested behavior. Such feedback may help participants confirm that their action was appropriate and meaningful.

Participants also suggested that the agents' intentions should be expressed more clearly. This included clearer motion, stronger gaze direction, and more direct indication of the target object or desired behavior. Some responses indicated that the horse agent's intention was difficult to interpret, possibly because participants had less prior knowledge about how to read horse-like emotional expressions or communicative behaviors. In the smartphone-use task, participants suggested more direct spatial interventions, such as approaching the user more closely or moving in front of the smartphone screen. These responses indicate that persuasive behaviors should be adjusted according to the task context and the spatial relationship between the user, the agent, and the target behavior.

**Table 8. Categorized free-description responses for suggested improvements.**

Values indicate the number of open-ended responses assigned to each improvement category. For example, a value of 6 in the U-cat column for Feedback in the feeding task means that six responses in that condition suggested adding clearer feedback after the participant performed the target behavior.

| Task | Category | U-dog | C-dog | B-dog | U-cat | C-cat | U-horse | C-horse | Total |
|---|---|---|---|---|---|---|---|---|---|
| Trash disposal | Feedback | 1 | 1 | 1 | 4 | 2 | 1 | 1 | 11 |
| Trash disposal | Clearer motion | 1 | 0 | 0 | 1 | 1 | 2 | 3 | 8 |
| Trash disposal | Physical/spatial intervention | 0 | 1 | 3 | 0 | 1 | 0 | 0 | 5 |
| Feeding | Feedback | 2 | 1 | 0 | 6 | 2 | 1 | 2 | 14 |
| Feeding | Clearer motion | 1 | 1 | 3 | 0 | 2 | 5 | 1 | 13 |
| Feeding | Physical/spatial intervention | 0 | 0 | 2 | 0 | 0 | 0 | 0 | 2 |
| Smartphone use | Physical/spatial intervention | 3 | 1 | 1 | 3 | 1 | 3 | 1 | 13 |
| Smartphone use | Clearer motion | 4 | 0 | 2 | 2 | 1 | 1 | 2 | 12 |
| Smartphone use | Feedback | 0 | 1 | 1 | 2 | 4 | 0 | 2 | 10 |

Overall, the qualitative findings complement the quantitative results by showing that participants relied on emotional expression, gaze direction, motion, and vocalization to interpret the agents' persuasive intentions. They also indicate that post-action feedback and clearer task-specific cues may improve persuasive effectiveness, especially for tasks that are difficult to communicate through spatial guidance alone.

## 5. Discussion

This study investigated whether persuasive behaviors using emotional expression and nonverbal motion can be generalized across quadruped virtual agents representing different animal species. The results provide three main findings. First, the bark-only condition yielded lower intention understanding and lower scores on several behavioral outcome measures in some tasks than conditions involving structured persuasive movements and emotional expressions. Importantly, the dog agent was visible in the bark-only condition; therefore, this result does not indicate that visual presentation of an animal-like agent was absent or ineffective. Rather, it suggests that the visual presence of an agent and simple barking alone are insufficient for clearly communicating persuasive intention. Second, no consistent significant differences were observed between species-specific and shared behavior conditions, suggesting that faithful reproduction of animal-specific motion was not the primary determinant of persuasive effectiveness in the present tasks. Third, psychological reactance and discomfort remained low across conditions, indicating that quadruped virtual agents may support behavior change without strongly eliciting negative emotional responses.

These findings should be interpreted in relation to the broader perspective of Persuasive and Affective Human–AI Interaction. The present study suggests that persuasive effectiveness in animal-like virtual agents depends not simply on animal appearance or species-specific realism, but on whether users can interpret the agent's intention, emotional state, and need for support through structured affective and nonverbal cues. In this sense, the comparison of dog, cat, and horse agents was not merely a comparison of different animal forms. Rather, it served as a methodological approach for examining whether persuasive behaviors can be abstracted as species-independent functional design elements.
The following sections discuss these findings in detail. Section 5.1 examines the role of emotional expression and intention readability. Section 5.2 discusses why persuasive effectiveness may depend more on functional cues than on species-specific motion. Section 5.3 considers persuasion through vulnerability, familiarity, and low psychological reactance. Section 5.4 discusses the role of presence and controllability in mixed reality. Section 5.5 derives cross-species design patterns for persuasive quadruped agents. Section 5.6 discusses broader implications for Persuasive and Affective Human–AI Interaction, and Section 5.7 summarizes limitations and future work.

### 5.1 Emotional Expression and Intention Readability in Quadruped Persuasive Agents

One of the clearest findings of this study is that the bark-only condition showed lower intention understanding than the conditions involving structured persuasive movements and emotional expressions, particularly in the trash disposal and feeding tasks. This result should be interpreted carefully. The bark-only condition was not an audio-only condition. The dog agent was visually present, but it performed only a barking animation without the full set of emotional expressions, gaze alternation, attention guidance, guiding movements, and pointing behaviors. Therefore, the lower scores in this condition suggest that the mere visual presence of an animal-like agent and simple barking alone are insufficient to communicate persuasive intention clearly.

This finding supports the view that emotional expression in pet-type persuasive agents functions as more than an affective decoration. Emotional expression can serve as a functional cue that helps users infer what the agent wants, what state the agent is in, and how the user may respond. In the present study, participants' free descriptions often referred to the agent's apparent emotional state, such as sadness, loneliness, hunger, or discomfort. These descriptions suggest that participants did not simply perceive the agent as a moving object, but interpreted it as an intentional and affective entity. Such interpretation appears to be important for translating the agent's behavior into users' own behavioral responses.

The results are consistent with previous findings showing that sadness and confusion in dog-like pet-type agents can motivate supportive behavior when users interpret these emotions as signs that the agent needs help (Harada and Sumi, 2024). They also extend subsequent work showing that persuasive effects depend on the combination of emotional expression, behavioral cues, and context in mixed reality (Sumi and Harada, 2025). In the present study, the persuasive behaviors were applied not only to a dog agent but also to cat and horse agents. The fact that structured movements and emotional expressions produced higher intention understanding and behavioral outcomes than the bark-only condition in several tasks suggests that intention readability can be supported by functional cues even when the agent is not human-like.

In addition to emotional expression, attention-directing behaviors also played an important role in intention readability. Behaviors such as looking at the user, moving toward a target, alternating gaze between an object and its destination, and returning attention to the user provided a structure through which participants could infer the intended action. These cues are particularly important for nonverbal or minimally verbal agents because users must infer the agent's intention from its behavior rather than from explicit instruction. In this sense, the persuasive behaviors used in this study can be understood as structured nonverbal communication, in which emotional expression, gaze direction, spatial movement, and object-directed behavior work together to make the agent's goal readable.

The task-dependent results also indicate that intention readability is not determined by the agent's behavior alone. In the trash disposal and feeding tasks, the target action could be visually and spatially linked to objects in the environment, such as trash, a trash box, food, or a feeding dish. This may have made gaze alternation, guiding behavior, and pointing easier to interpret. In contrast, the smartphone-use task was less directly tied to a physical destination or object transformation, and the intention may have been more difficult to infer through the same behavioral structure. This suggests that persuasive behaviors should be designed in relation to the target task and the environmental affordances available in the interaction.

Overall, these findings suggest that intention readability is a central design factor in quadruped persuasive agents. A visible animal-like agent may attract attention and create affinity, but persuasive effectiveness requires more than presence or simple vocalization. It depends on whether users can interpret the agent's emotional state, attentional focus, and behavioral goal through coordinated affective and nonverbal cues. Thus, emotional expression and attention guidance should be regarded as core functional design elements for persuasive quadruped virtual agents.

### 5.2 Beyond Species-Specific Motion: Functional Cues for Persuasion

A central question in this study was whether persuasive effectiveness in quadruped virtual agents depends on species-specific expressive motion. The results did not support this assumption. Across the tasks and evaluation measures, no consistent significant advantage was observed for species-specific behavior over shared behavior. This suggests that, at least in the present experimental tasks, persuasive effectiveness was not primarily determined by whether the agent's movements were adapted to dog-like, cat-like, or horse-like expressive styles.

This finding is important because animal-like agent design often assumes that greater species-specific motion design will lead to greater naturalness, familiarity, or effectiveness. Such realism may be valuable for improving believability or aesthetic quality. However, the present results suggest that species-specific motion design is not necessarily the key factor for persuasive behavior. Instead, what appears to matter more is whether users can read the agent's intention, emotional state, and behavioral goal. In other words, persuasive effectiveness may depend less on whether the movement is biologically accurate and more on whether the movement functions as a readable cue for the user.

The comparison between species-specific and shared behavior conditions provides initial evidence for this interpretation. Shared behaviors, such as looking at the user, moving toward a target, alternating gaze between relevant objects, and returning attention to the user, were not tied to a single animal species. Nevertheless, they could still support intention understanding and behavioral response. These behaviors can be understood as functional cues because they guide the user's attention, indicate a target, and make the agent's goal interpretable. From this perspective, the effectiveness of persuasive behavior lies not in reproducing a particular animal's motion, but in organizing cues so that users can infer what the agent wants them to do.

This interpretation does not mean that species-specific motion is unimportant. Species-specific behaviors may still contribute to naturalness, enjoyment, familiarity, or long-term attachment. They may also be important in contexts where users expect a high level of realism or where the animal species itself carries strong symbolic or relational meaning. However, the present findings suggest that species-specific motion was not a necessary condition for persuasive effectiveness in the everyday tasks examined here. Rather, persuasive interaction could be supported by more general design elements that operate across animal species.

This result also helps clarify the role of cross-species comparison in the present study. The purpose of comparing dog, cat, and horse agents was not simply to identify which animal was most persuasive. Rather, species variation was used as a methodological tool to examine whether persuasive effects depend on animal-specific familiarity and motion or on more abstract functional cues. The absence of consistent differences between species-specific and shared behavior conditions suggests that persuasive mechanisms can be partially abstracted from animal-specific motion and described in terms of functional design elements, such as emotional expression, attention guidance, gaze alternation, and pointing, that support intention readability.

At the same time, the results should be interpreted with caution. The present study examined three quadruped species and three everyday tasks in a controlled mixed reality setting. It is possible that species-specific motion would become more important in situations where users have stronger expectations about how a particular animal should move, or where the task requires more complex or characteristic animal behaviors. For example, clearer effects of species-specific motion may emerge with other animal species, longer-term interactions, or tasks in which animal-specific movement styles are more directly related to the user's interpretation of the agent's intention. Therefore, the present findings should be understood as initial evidence for species-independent persuasive design, rather than as evidence that species-specific motion is generally unnecessary.

Taken together, these findings suggest that a central design factor for persuasion in quadruped virtual agents is not species-specific motion fidelity itself, but emotion-mediated intention readability. In other words, persuasive effectiveness depends on whether users can interpret the agent's affective state, attentional focus, and intended action from its behavior. Animal-like appearance and species-appropriate motion may support this interpretation, but they appear to function primarily as expressive forms through which emotional state, need, and communicative intention become readable.

Overall, this study suggests that persuasive quadruped agents can be designed at a functional level. Instead of beginning with the question of how to reproduce each animal's behavior as accurately as possible, designers may first ask what communicative function the behavior should serve: attracting attention, expressing need, guiding the user, indicating a target, or, in future designs, confirming an action. Once these functions are defined, they can be implemented in ways that are appropriate to each animal species while preserving a shared persuasive structure. This approach provides a basis for standardizing persuasive behaviors across animal-like virtual agents.

### 5.3 Persuasion Through Vulnerability, Familiarity, and Low Reactance

Another important finding of this study is that psychological reactance and discomfort remained low across all conditions. This suggests that the persuasive behaviors used in the present study did not strongly elicit negative emotional responses, even though the agents attempted to influence participants' everyday actions. This is important for persuasive technology because behavior change systems must not only promote target behaviors, but also avoid being perceived as overly controlling, intrusive, or burdensome. This interpretation is also consistent with later work on psychological reactance, which conceptualizes reactance as involving both negative affect and resistance-related cognition when persuasive attempts are perceived as threatening users' freedom (Dillard and Shen, 2005).

The low levels of reactance and discomfort may be related to the nature of animal-like persuasive agents. Unlike humanoid agents or explicit recommendation systems, pet-type agents do not necessarily persuade by giving commands, explanations, or arguments. Instead, they can invite users to act by appearing to need support, attention, or care. In the present study, participants' free descriptions often referred to the agents as looking sad, lonely, hungry, or troubled. These responses suggest that participants may have interpreted the agents' behaviors not as demands imposed by a system, but as cues indicating that the agent needed help or wished to communicate something.

This interpretation is consistent with the concept of weak robots, in which apparent vulnerability, incompleteness, or dependence on others can elicit human assistance and create relational forms of interaction (Okada, 2023). It is also consistent with studies on help-eliciting robots, such as the sociable trash box, which showed that robots can encourage human assistance not by commanding people, but by displaying limited capability or neediness (Yamaji et al., 2011). In the context of the present study, sadness, hesitation, gaze alternation, and approach behavior may have functioned as vulnerability cues that encouraged participants to support the agent rather than simply comply with an instruction.

This mechanism helps explain why quadruped virtual agents may be suitable for subtle behavior change. When an animal-like agent appears to need attention or support, users may interpret the target behavior as a voluntary supportive action rather than as compliance with an external demand. In this sense, pet-type persuasive agents may reduce psychological burden by reframing persuasion as care, assistance, or social response rather than as obedience to an instruction.

The relationship between familiarity and actual behavior further supports this relational interpretation. Familiarity with the animal species was positively associated with actual behavior in some conditions, especially in the cat and horse conditions and in the dog bark-only condition. This exploratory result suggests that when behavioral cues are less easily interpreted or when users have less shared knowledge about the animal species, users' prior affinity toward the animal may play a larger role in motivating action. For cat and horse agents, individual differences in familiarity may have influenced whether participants responded behaviorally to the agent. This does not necessarily mean that familiarity directly changed how participants interpreted the agent's intention, but it suggests that prior affinity toward the animal may have influenced the likelihood of acting on the agent's cues.

In contrast, familiarity did not show the same pattern for the dog agents with structured persuasive behaviors. One possible explanation is that dogs are generally familiar social animals, and many users may already share expectations about canine communication. Therefore, structured persuasive behaviors by the dog agent may have been relatively easy to interpret regardless of individual differences in familiarity. However, this interpretation should be treated as exploratory, because the present study did not directly measure users' prior knowledge of animal communication.

These findings indicate that persuasive effectiveness in quadruped virtual agents is shaped not only by the design of expressive behaviors, but also by the user's relationship to the agent. Emotional expression and nonverbal cues can communicate intention, but their effects may be influenced by familiarity, affinity, and perceived relationship. Therefore, designing persuasive animal-like agents requires attention not only to what the agent does, but also to how users are likely to perceive and relate to the agent.

Overall, the present findings suggest that animal-like virtual agents can support a form of low-reactance persuasion based on vulnerability, familiarity, and relational engagement. These agents may be useful for everyday behavior change contexts where designers wish to encourage action without increasing users' sense of being pressured or controlled. Rather than relying on direct commands, such agents may support behavior change by inviting users to interpret the agent as needing attention, care, or support.

**5.4 Presence and Controllability in MR-Based Animal-Like Agents**

The present study also highlights the importance of mixed reality as a platform for persuasive animal-like agents. In the experiment, the agents did not have physical bodies, but they were spatially presented in the participants' real environment and directed their behaviors toward real objects and locations. This means that the persuasive interaction was not merely a video-based presentation of animal motion. Rather, the agents were presented as situated entities that appeared in the same physical space as the participants and acted in relation to the surrounding environment.

This point is important because physical embodiment is not the only way to create a sense of presence. In mixed reality, virtual agents can be anchored in real space and can direct gaze, movement, emotional expression, and attention toward real-world objects. Such spatial grounding can make the agent's behavior more meaningful because users can interpret the agent's actions in relation to their own environment. For example, when the agent alternates its gaze between an object and its destination, or moves toward a target location, the behavior becomes interpretable as an attempt to guide the user's attention or communicate an intended action.

The results of the present study suggest that this situatedness contributed to persuasive effectiveness. In the trash disposal and feeding tasks, the target action was directly connected to physical objects in the environment, such as trash, a trash box, food, and a feeding dish. These objects provided clear spatial anchors for gaze alternation, guiding behavior, and pointing. In contrast, the smartphone-use task was less clearly tied to object movement or a spatial destination, and the same behavioral structure may have been less effective for communicating intention. This task-dependent difference suggests that MR-based persuasive agents may be particularly effective when their behaviors can be grounded in visible objects, places, and action possibilities in the user's physical environment.

At the same time, MR provides a level of controllability that is difficult to achieve with real animals and sometimes difficult even with physical robots. The designer can systematically manipulate the agent's species, motion, emotional expression, timing, position, and relationship to environmental objects while keeping the interaction context relatively consistent. This controllability is important not only for application design but also for research. It allows researchers to separate animal-specific appearance and motion from more general functional cues, such as emotional expression, attention guidance, pointing, and intention readability. In this sense, MR provides a useful platform for experimentally decomposing the design elements of persuasive interaction.

This controllability is especially relevant to the cross-species design perspective of the present study. Real animals differ in unpredictable ways, and physical robots often require different hardware designs for different animal forms. In contrast, MR-based virtual agents allow multiple animal species to be implemented within a common interaction framework. This makes it possible to compare dog, cat, and horse agents while controlling the structure of their persuasive behaviors. The absence of consistent differences between species-specific and shared behavior conditions can therefore be interpreted not only as a practical design finding, but also as evidence that MR is useful for examining which aspects of persuasive behavior are species-dependent and which are functionally generalizable. The findings also suggest that MR-based animal-like agents may combine a sense of co-presence with low psychological burden. Participants did not report high levels of discomfort or reactance, even though the agents attempted to influence their actions. One possible interpretation is that the agents were sufficiently situated to be interpreted as intentional and affective entities, while remaining non-intrusive and unlikely to be perceived as coercive. In other words, MR may provide a balanced form of presence: the agent can appear co-present and socially meaningful while remaining flexible, controllable, and non-invasive.

Overall, the present study positions MR as a promising medium for persuasive and affective Human–AI Interaction. MR-based animal-like agents can make use of real-world context, spatial anchoring, and social presence while allowing precise control over expressive and behavioral cues. This combination is particularly valuable for designing subtle behavior change interventions, because it enables agents to guide attention and communicate intention through situated nonverbal behavior rather than through direct verbal instruction.

**5.5 Toward Cross-Species Design Patterns for Persuasive Quadruped Agents**

The present findings provide initial implications for designing persuasive behaviors that can be reused across quadruped animal-like agents. The goal of standardization in this context is not to make all animals move in exactly the same way, nor to ignore the unique characteristics of each animal species. Rather, standardization should be understood as the identification of functional design elements that support intention readability, affective interpretation, and behavioral response across different animal forms.

The first design element is emotional expression. The results suggest that emotional expression helps users interpret the agent's state and infer why the agent is acting. Expressions such as sadness, loneliness, hunger, or discomfort can make the agent's need more readable and can encourage users to respond supportively. However, emotional expression should not be treated as a general enhancer of persuasion. As shown in prior work and supported by the present findings, the effectiveness of an emotion depends on whether it is consistent with the target behavior and the interaction context. Therefore, designers should select emotional expressions based on the behavioral goal. For example, vulnerability-related expressions may be suitable when the target behavior can be interpreted as helping the agent, whereas expressions that suggest playfulness or anger may lead to different interpretations.

The second design element is attention calling. Before an agent can persuade users, it must first attract their attention and establish that it is attempting to communicate. In the present study, looking at the user, vocalizing, approaching, and returning attention to the user functioned as cues that helped participants notice the agent and interpret its behavior as directed toward them. This is particularly important for nonverbal agents, because the beginning of the interaction must signal that the agent's behavior is meaningful rather than random movement.

The third design element is gaze alternation and pointing. The findings suggest that alternating gaze or head orientation between the user, the target object, and the destination can make the agent's intention more readable. This appeared particularly relevant in tasks such as trash disposal and feeding, where the target behavior could be spatially grounded in visible objects. Gaze alternation and pointing can therefore be treated as cross-species functional cues for indicating "what the agent is referring to" and "what action the user is expected to infer." These cues may be implemented differently depending on the animal species, but their communicative function can remain shared.

The fourth design element is guiding behavior. Moving toward a target location or object can help users understand the direction of the intended action. Guiding is particularly useful when the target behavior involves a spatial relationship, such as moving an object to a destination or approaching a specific location. In quadruped agents, guiding can be designed as a natural-looking movement while still preserving a common persuasive structure across species.

The fifth design element is post-action feedback. Although the present experiment mainly focused on the persuasion phase, participants' free descriptions suggested that feedback after the user's action would make the interaction more understandable and satisfying. For example, showing that the agent is relieved, happy, calm, or satisfied after the user acts may help users understand that their behavior was appropriate and meaningful. This feedback could also support continued engagement and may be important for longer-term behavior change.

These design elements suggest that persuasive behavior in quadruped agents can be organized as a sequence of functional cues: attracting the user's attention, expressing need or affective state, indicating the target object or location, guiding the user's interpretation of the desired behavior, and responding after the user acts. This sequence does not need to be identical across animal species at the level of motion. Instead, each species can express the same communicative function through movements that are appropriate to its body form and social image.

This view also clarifies the role of species-specific design. Species-specific motion may be useful for increasing naturalness, believability, or user attachment, but the present results suggest that it should be built around functional persuasive cues rather than treated as the primary basis of persuasion. In other words, designers may first define the persuasive function, such as calling attention, expressing vulnerability, pointing to an object, or confirming a completed action, and then adapt the surface motion to each animal species.

The proposed design pattern is therefore species-independent at the functional level and species-sensitive at the expressive level. This distinction is important for building reusable behavior patterns. If persuasive behaviors are standardized only as fixed animations, they may fail when applied to animals with different body structures or user expectations. However, if they are standardized as functional units, they can be adapted to different quadruped agents while preserving their persuasive purpose.

The size control used in this study also has practical implications for cross-species design. In real-world applications, animal-like agents are not necessarily presented at their biologically realistic size. For example, a horse-like character may be scaled down for domestic, tabletop, or small-room interaction contexts, whereas a cat-like or dog-like character may be scaled to maintain visibility and readability in an MR environment. Therefore, the present study did not treat realistic body size as a primary design factor. Instead, the agents were adjusted to a comparable apparent size so that the experiment could focus on functional persuasive cues, such as emotional expression, attention calling, gaze alternation, pointing, and guiding. This approach is consistent with practical character-based agent design, where animal species can be used as expressive forms while communicative functions are designed independently of biological scale.

Overall, the present study suggests that cross-species design patterns for persuasive quadruped agents should be based on functional cues rather than species-specific motion design alone. Emotional expression, attention calling, gaze alternation, pointing, guiding, and post-action feedback can serve as reusable design elements for animal-like persuasive agents. Such design patterns may support the development of scalable, reusable, and socially acceptable persuasive behaviors for non-humanoid agents in mixed reality and other interactive environments.

### 5.6 Implications for Persuasive and Affective Human–AI Interaction

The findings of this study have broader implications for Persuasive and Affective Human–AI Interaction. The present results suggest that non-humanoid AI agents can support behavior change through affective, embodied, and situated interaction, even without human-like appearance, language, or explicit instruction. In this sense, quadruped virtual agents expand the design space of persuasive agents beyond humanoid or conversational systems. They suggest that persuasion can emerge from the user's interpretation of emotional expression, bodily movement, spatial behavior, and the agent's apparent need or intention.

This perspective is important for PAHAI because it emphasizes persuasion as an affective and relational process rather than as a purely informational or directive process. In the present study, participants often responded to the agents by interpreting their emotional states, such as sadness, loneliness, hunger, or discomfort. These interpretations were closely related to the participants' behavioral responses. This suggests that persuasive interaction can be designed not only by optimizing messages or recommendations, but also by shaping how users emotionally and socially interpret an agent's behavior.

A particularly important implication is that behavior change may occur through users' affective interpretation of an artificial entity. The agent does not need to possess real emotions in order to influence human behavior. Rather, what matters for persuasion is that users can perceive and imagine the agent as having an emotional state, intention, or need in a given situation. Even when users understand that the agent is artificial, they may still interpret it as sad, lonely, hungry, troubled, or in need of support. This imagined affective state can motivate users to act supportively. Thus, one possible persuasive pathway does not depend on whether the artificial agent truly has emotions, but on whether users attribute emotion, need, or vulnerability to the agent and translate that interpretation into action.

This interpretation is consistent with theories of anthropomorphism and mind perception, which suggest that people may attribute intentions, emotions, and experiential states to nonhuman agents when those agents provide sufficient social or behavioral cues (Epley et al., 2007; Gray et al., 2007). The present findings extend this view by showing that such affective attribution may not only shape users' impressions of an artificial agent, but may also contribute to behavior change.

The results also suggest that PAHAI should include non-humanoid and animal-like agents as important design targets. Humanoid agents can use speech, facial expression, and human-like gestures to communicate persuasive intentions. However, animal-like agents rely more strongly on nonverbal and affective cues, such as gaze direction, approach behavior, vocalization, and bodily orientation. This creates a different form of human–AI relationship, one in which users may respond through care, support, and empathy rather than through compliance with a recommendation. Such interaction may be particularly valuable for behavior change contexts in which subtle, low-pressure, and socially acceptable intervention is desirable.

The present findings are also relevant to the design of behavior change technologies in everyday environments. Many behavior change systems aim to support daily habits, such as cleaning, reducing excessive smartphone use, eating, exercising, studying, or maintaining routines. In these contexts, direct reminders or commands may be effective in some cases, but they may also become repetitive, annoying, or easy to ignore. Animal-like persuasive agents offer an alternative approach: they can embed behavioral prompts in affective and relational interaction. For example, an agent can express need, guide attention to an object, or show relief after the user acts. Such interactions may make behavior change feel less like external control and more like a meaningful social response.

Mixed reality further strengthens this possibility by allowing agents to intervene in relation to the user's physical environment. MR-based persuasive agents can be placed near real objects, look toward them, move around them, and make the intended action visible within the user's actual context. This makes MR particularly suitable for situated behavior change, where the goal is not merely to persuade users in abstract terms, but to guide attention and action in the moment and place where the behavior can occur. In this respect, MR-based animal-like agents can function as affective interfaces between users, real-world objects, and desired actions.

These implications also extend to education, healthcare, and affective learning. In educational contexts, animal-like agents could encourage learners to engage with tasks, attend to learning materials, or reflect on their actions through emotional and relational cues. In healthcare and well-being contexts, they could support low-burden prompts for daily routines, rehabilitation, medication adherence, or stress reduction. In affective learning and serious games, such agents could be used to create emotionally meaningful interactions that support engagement, reflection, and behavioral change. Although the present study did not directly evaluate learning or long-term health outcomes, it provides initial design evidence that affective and nonverbal interaction can shape users' interpretation and immediate action.

At the theoretical level, this study suggests that PAHAI should attend to three layers of design. The first is the perceptual layer, which concerns whether users can perceive the agent's behavior, emotional expression, and spatial orientation. The second is the interpretive layer, which concerns whether users can infer or imagine the agent's intention, need, or emotional state. This layer is particularly important because users may change their behavior not simply in response to what an artificial agent objectively does, but in response to what they believe or imagine the agent is feeling or trying to communicate. Even when users know that the agent is an artificial object, they may interpret it as sad, lonely, hungry, troubled, or in need of support, and this affective interpretation can motivate supportive behavior. The third is the behavioral layer, which concerns whether users translate this interpretation into action. The present findings suggest that persuasive effectiveness depends on the alignment of these layers. An agent may be visually present, but if its intention or affective state is not readable, persuasion may fail. Conversely, when emotional expression, attention guidance, and spatial behavior allow users to imagine the agent's need or emotional state in relation to the task, users may be more likely to respond behaviorally.

Overall, the present study contributes to PAHAI by showing that persuasive behavior can be designed through functional affective cues in non-humanoid virtual agents. Quadruped animal-like agents do not need to imitate humans to be persuasive. Instead, they can support behavior change by making their needs, intentions, and emotional states readable, or at least imaginable, through embodied and situated interaction. This finding provides a foundation for designing future AI agents that are not only intelligent or conversational, but also affective, relational, and gently persuasive in everyday life.

### 5.7 Limitations and Future Work

This study has several limitations. First, the number of participants was limited. Although the results provide initial evidence that emotional expression and structured nonverbal behaviors can support persuasive interaction in quadruped virtual agents, the sample size may have limited the statistical power of the analyses. Some behavioral measures showed clear tendencies but did not reach statistical significance in multiple comparisons. Future studies should include larger and more diverse participant samples to examine the robustness and generalizability of the findings.

Second, the present study examined only three quadruped animal species: dog, cat, and horse. These animals were selected because they differ in social familiarity, body size, and common human interpretations, while still sharing a quadruped body structure. However, it remains unclear whether the same design principles can be applied to other animal-like agents, such as birds, rabbits, wild animals, fantasy creatures, or more abstract non-humanoid agents. Future work should examine whether the functional cues identified in this study, such as emotional expression, gaze alternation, attention calling, guiding, pointing, and feedback, can be generalized to a wider range of agent forms. Relatedly, the apparent display size of the agents was controlled to be comparable across species. This control was appropriate for focusing on cross-species functional design patterns and for reflecting practical character-based agent design, but it means that the present study did not examine how biologically realistic or intentionally exaggerated body scale might influence presence, familiarity, perceived agency, or persuasive effectiveness.

Third, the bark-only condition was implemented only with the dog agent. This condition was useful as a baseline for examining whether the visual presence of a dog agent and simple barking alone were sufficient to communicate persuasive intention. However, because this baseline was not implemented for the cat and horse agents, it does not allow direct comparison of bark-only or vocalization-only effects across animal species. Future studies should design species-appropriate minimal-expression baseline conditions, such as meowing-only or neighing-only conditions, to examine how simple vocal and bodily cues differ across species.

Fourth, the tasks used in this study were short-term everyday actions conducted in a controlled experimental setting. The tasks, including trash disposal, feeding, and refraining from smartphone use, were suitable for examining immediate intention understanding and behavioral response. However, they do not directly demonstrate long-term behavior change, habit formation, or sustained interaction with the agent. Future research should investigate whether quadruped persuasive agents can support repeated or long-term behavior change in everyday environments.

Fifth, the effectiveness of the persuasive behaviors appeared to depend on task characteristics. The trash disposal and feeding tasks were spatially grounded in visible objects, whereas the smartphone-use task was less directly connected to object movement or a clear destination. This suggests that the effectiveness of emotional expression, gaze alternation, and guiding behavior may depend on whether the target behavior can be visually and spatially represented in the environment. Future studies should compare a wider range of tasks with different levels of spatial grounding, behavioral cost, social meaning, and personal relevance.

Sixth, individual differences were not fully examined. The present results suggest that familiarity with the animal may influence actual behavior in some conditions. However, the study did not systematically measure participants' prior experiences with animals, preferences, cultural background, personality, empathy, or attitudes toward artificial agents. These factors may affect how users interpret the agent's emotional expression, need, or intention. Future work should examine how individual and cultural differences moderate persuasive effects in animal-like virtual agents.

Seventh, the present study focused mainly on predesigned behaviors and did not examine adaptive interaction. In real-world persuasive Human–AI Interaction, agents may need to adjust their emotional expression, timing, distance, intensity, or feedback based on the user's responses. Future studies should explore adaptive quadruped agents that can modify their behavior according to user behavior, affective state, familiarity, and task context. Such adaptive interaction would be particularly important for developing personalized persuasive and affective AI agents.

Finally, the present study examined users' interpretation and behavior, but did not directly measure the cognitive and affective process through which users imagined the agent's emotional state. The discussion suggests that users may change their behavior by attributing emotion, need, or vulnerability to an artificial agent. However, future research should more directly investigate this interpretive process through interviews, detailed behavioral coding, physiological measures, or questionnaires on perceived emotion, agency, and intentionality. Such work would deepen the theoretical understanding of how affective interpretation of artificial agents contributes to behavior change.

Despite these limitations, the present study provides initial evidence that persuasive behaviors in quadruped virtual agents can be designed at the level of species-independent functional cues. The findings suggest that emotional expression, attention guidance, gaze alternation, guiding behavior, and post-action feedback may serve as reusable design elements for animal-like persuasive agents. Future work should extend these findings to longer-term, adaptive, and real-world applications, thereby contributing to the development of Persuasive and Affective Human–AI Interaction.

## 6. Conclusion

This study investigated persuasive behaviors in quadruped virtual agents from a cross-species design perspective. Specifically, we examined whether emotional expression and nonverbal persuasive movements can support behavior change across dog, cat, and horse agents, and whether species-specific motion enhances persuasive effectiveness compared with shared behavior across species.

The results showed that the bark-only condition produced lower intention understanding and lower scores on several behavioral outcome measures in some tasks than conditions involving structured persuasive movements and emotional expressions. Importantly, this condition did not remove the visual presence of the animal-like agent. Rather, it showed that a visible dog agent and simple barking alone were not sufficient to clearly communicate persuasive intention. These findings indicate that emotional expression, gaze alternation, attention guidance, pointing, and guiding behavior play important roles in making the agent's intention readable and in supporting users' behavioral responses.

In contrast, no consistent significant differences were observed between species-specific behavior and shared behavior conditions. This suggests that persuasive effectiveness in the present tasks was not primarily determined by species-specific expressive motion design. Instead, the findings indicate that persuasion in quadruped virtual agents may rely more on species-independent functional cues, such as emotional expression, attention guidance, spatial grounding, and intention readability. This provides initial evidence that persuasive behaviors in animal-like agents can be abstracted and standardized at the level of communicative function, while still allowing expressive adaptation to each animal species.

The study also showed that psychological reactance and discomfort remained low across conditions. This suggests that quadruped animal-like agents may support a gentle form of behavior change that does not strongly elicit negative emotional responses. Furthermore, familiarity with the animal species was associated with actual behavior in some conditions, indicating that users' prior affinity toward a particular animal may influence whether they respond to its persuasive cues.

Overall, this study contributes to Persuasive and Affective Human–AI Interaction by showing that non-humanoid virtual agents can influence human behavior through affective, embodied, and situated cues. The findings suggest that users may change their behavior not simply because an artificial agent provides information or commands, but because they interpret or imagine the agent's emotional state, intention, or need for support. In this sense, one possible persuasive pathway lies not in whether the agent truly has emotions, but in whether users can attribute emotion, need, or vulnerability to the agent and translate that interpretation into action.

These findings provide design implications for developing reusable behavior patterns for persuasive quadruped virtual agents. Emotional expression, attention calling, gaze alternation, pointing, guiding, and post-action feedback can serve as functional design elements that may be applied across animal-like agents. Future research should examine larger and more diverse participant samples, additional animal species and agent forms, species-appropriate baseline conditions, long-term behavior change, adaptive interaction, and individual or cultural differences in interpreting animal-like agents. By extending these directions, quadruped virtual agents may become a useful platform for subtle, affective, and socially acceptable behavior change in mixed reality and everyday Human–AI Interaction.

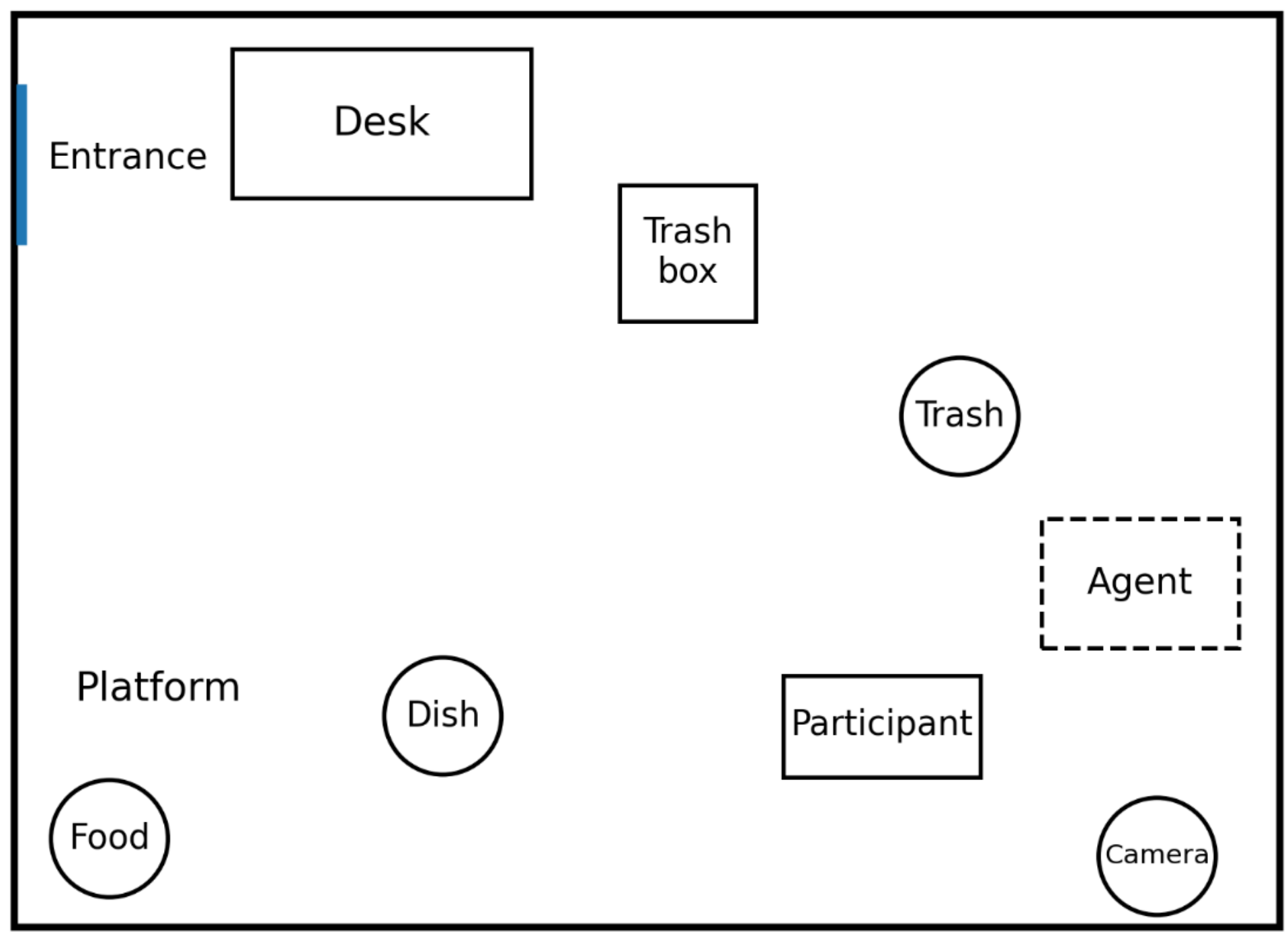

**Supplementary Figure S1. Spatial arrangement of the experimental room and task-related objects.**

The diagram shows the arrangement of the participant, the initial position of the virtual agent, the task-related objects, and the recording device used in the MR experiment. The trash disposal task involved the trash and trash box, the feeding task involved the food and feeding dish, and the smartphone-use task involved the participant's smartphone. The experimenter operated the control application from outside the experimental room, and the physical objects were reset to their initial positions before each condition.